\documentclass{article}
\PassOptionsToPackage{numbers, compress}{natbib}
 \usepackage[eandd,preprint]{neurips_2026}

\usepackage{amsmath,amssymb}
\usepackage{graphicx}
\usepackage{booktabs}
\usepackage{hyperref}
\usepackage[T1]{fontenc}
\usepackage{microtype}
\usepackage{subcaption}
\usepackage{listings}
\usepackage{xcolor}
\usepackage[scale=0.95,semibold]{sourcecodepro}
\usepackage{natbib}
\usepackage{tikz}
\usetikzlibrary{arrows.meta,positioning,calc,decorations.pathmorphing,backgrounds}
\usepackage[most]{tcolorbox}
\usepackage{enumitem}
\usepackage{wrapfig}
\usepackage{needspace}

\makeatletter
\renewcommand{\paragraph}{%
  \@startsection{paragraph}{4}{\z@}%
                {0.4ex \@plus 0.2ex \@minus 0.1ex}%
                {-1em}%
                {\normalsize\bf}%
}
\makeatother

\newtcolorbox{principlesbox}{
  colback=blue!2,
  colframe=blue!40!black,
  boxrule=0.4pt,
  arc=1.5pt,
  left=6pt, right=6pt, top=4pt, bottom=4pt,
}
\newtcolorbox{slopbox}[1][]{
  colback=red!3,
  colframe=red!40!black,
  title={#1},
  fonttitle=\bfseries\small,
  boxrule=0.4pt,
  arc=1.5pt,
  left=2pt, right=2pt, top=2pt, bottom=2pt,
}
\newtcolorbox{cleanbox}[1][]{
  colback=green!3,
  colframe=green!40!black,
  title={#1},
  fonttitle=\bfseries\small,
  boxrule=0.4pt,
  arc=1.5pt,
  left=2pt, right=2pt, top=2pt, bottom=2pt,
}
\definecolor{runexborder}{RGB}{0,32,96}
\newtcolorbox{runningexample}[1][Running Example]{
  enhanced, breakable,
  colback=white,
  colframe=runexborder,
  boxrule=1pt,
  sharp corners,
  fonttitle=\bfseries,
  coltitle=runexborder,
  colbacktitle=white,
  title=#1,
  attach boxed title to top left={xshift=10pt, yshift=-\tcboxedtitleheight/2},
  boxed title style={
    colback=white, colframe=white,
    boxrule=0pt, sharp corners,
    left=4pt, right=4pt, top=0pt, bottom=0pt,
  },
  top=10pt,
  left=10pt, right=4pt, bottom=4pt,
}

\definecolor{diffadd}{HTML}{dafbe1}
\definecolor{diffrem}{HTML}{ffcecb}
\definecolor{diffaddmark}{HTML}{1a7f37}
\definecolor{diffremmark}{HTML}{cf222e}
\definecolor{diffctx}{HTML}{f6f8fa}
\definecolor{diffborder}{HTML}{d0d7de}

\usepackage{colortbl}
\usepackage{array}
\usepackage{multirow}
\usepackage{longtable}
\usepackage{tabularx}

\newtcolorbox{diffbox}[1][]{
  colback=white,
  colframe=diffborder,
  title={#1},
  fonttitle=\bfseries\small,
  boxrule=0.5pt,
  arc=2pt,
  left=0pt, right=0pt, top=0pt, bottom=0pt,
  boxsep=0pt,
}

\definecolor{figyellow}{HTML}{FFD966}
\definecolor{figred}{HTML}{E06666}
\definecolor{codebg}{gray}{0.97}
\definecolor{codeframe}{gray}{0.75}
\definecolor{codegreen}{rgb}{0.0,0.4,0.0}
\definecolor{codepurple}{rgb}{0.5,0.0,0.33}
\definecolor{codenumber}{gray}{0.55}

\lstdefinestyle{specblock}{
  basicstyle=\ttfamily\footnotesize,
  numbers=none,
  xleftmargin=1em,
  language={},
  keywordstyle={},
  stringstyle={},
  commentstyle={},
}

\title{SlopCodeBench: Benchmarking How Coding Agents Degrade Over Long-Horizon Iterative Tasks}

\author{
  Gabriel Orlanski\textsuperscript{1}\thanks{Corresponding author. Correspondence to \texttt{gorlanski@wisc.edu}.} \And
  Devjeet Roy\textsuperscript{2} \And
  Alexander Yun\textsuperscript{1} \And
  Changho Shin\textsuperscript{1} \And
  Alex Gu\textsuperscript{3} \And
  Albert Ge\textsuperscript{1} \And
  Dyah Adila\textsuperscript{1} \And
  Nicholas Roberts\textsuperscript{1} \And
  Frederic Sala\textsuperscript{1} \And
  Aws Albarghouthi\textsuperscript{1} \AND
  \normalfont\textsuperscript{1}University of Wisconsin--Madison \quad
  \textsuperscript{2}Washington State University \quad
  \textsuperscript{3}MIT
}

\makeatletter
\if@anonymous
  \newcommand{\scbsiteurl}{https://anonymous.4open.science/r/anon-slop-code-CA9D/}
  \newcommand{\scbrepourl}{https://anonymous.4open.science/r/anon-slop-code-CA9D/}
  \newcommand{\scbproblemurl}[1]{anonymous.4open.science/r/anon-slop-code-CA9D/}

\else
  \newcommand{\scbsiteurl}{https://www.scbench.ai}
  \newcommand{\scbrepourl}{https://github.com/gabeorlanski/scb-problems}
  \newcommand{\scbproblemurl}[1]{https://www.scbench.ai/problems/#1}

\fi
\makeatother

\newcommand{\scb}{SlopCodeBench}
\newcommand{\problem}{P}

\newcommand{\ckpt}[1]{C_{#1}}
\newcommand{\ckptinit}{\ckpt{1}}
\newcommand{\solution}[1]{y_{#1}}
\newcommand{\solutioninit}{\solution{0}}

\newcommand{\spec}[1]{x_{#1}}
\newcommand{\agent}{\pi_{\theta}}
\newcommand{\codesearch}{\texttt{code\_search}}

\newcommand{\justsolve}{\texttt{just-solve}}
\newcommand{\antislop}{\texttt{anti-slop}}
\newcommand{\planfirst}{\texttt{plan-first}}

\newcommand{\overallNumModels}{15}
\newcommand{\overallTotalCheckpoints}{196}
\newcommand{\overallBestModel}{GPT~5.5}
\newcommand{\overallBestStrictSolveRate}{14.8}

\newcommand{\overallIsoSolveMax}{28.1}

\newcommand{\overallCoreSolveMax}{67.3}
\newcommand{\overallChurnMeanEarlyPct}{97.4}
\newcommand{\overallChurnMeanLatePct}{29.5}
\newcommand{\overallTokensTotalB}{13.18}

\newcommand{\overallTokenLowestModel}{Composer~2}
\newcommand{\overallTokenLowestNormB}{0.38}
\newcommand{\overallTokenHighestModel}{Kimi~K2.6}
\newcommand{\overallTokenHighestNormB}{1.14}

\newcommand{\overallTestCorePassEarlyPct}{64.6}
\newcommand{\overallTestCorePassLatePct}{35.5}
\newcommand{\overallTestFuncPassEarlyPct}{62.7}
\newcommand{\overallTestFuncPassLatePct}{57.7}
\newcommand{\overallTestErrorPassEarlyPct}{80.1}
\newcommand{\overallTestErrorPassLatePct}{62.2}

\newcommand{\transProblems}{36}

\newcommand{\erosionPctIncreasing}{77}
\newcommand{\erosionHighCcCountFirstMean}{3.6}
\newcommand{\erosionHighCcCountLastMean}{23.7}
\newcommand{\erosionForgeNumCkpts}{8}
\newcommand{\erosionForgeMainCcFirst}{13}
\newcommand{\erosionForgeMainCcLast}{92}
\newcommand{\erosionForgeMainCcMultiplier}{7}
\newcommand{\erosionForgeMainLinesFirst}{38}
\newcommand{\erosionForgeMainLinesLast}{240}

\newcommand{\vfTrajIncPct}{75.5}
\newcommand{\vfCloneGrowthPct}{96}
\newcommand{\vfAstOverallGrowth}{0.3}

\newcommand{\abFigCoreIsoRatioMin}{2.5}
\newcommand{\abFigCoreIsoRatioMax}{5.4}
\newcommand{\abFigCostGrowth}{2.2}

\newcommand{\newErosionFirstCcMaxMean}{27.5}
\newcommand{\newErosionLastCcMaxMean}{69.0}

\newcommand{\avhPanelN}{473}
\newcommand{\avhPanelStartYear}{1990}
\newcommand{\avhPanelEndYear}{2026}

\newcommand{\avhAgentVerbMean}{0.44}
\newcommand{\avhAllVerbRatio}{2.3}
\newcommand{\avhAllErosionRatio}{2.0}
\newcommand{\avhHobbyN}{114}
\newcommand{\avhNicheN}{124}
\newcommand{\avhEstabN}{120}
\newcommand{\avhMajorN}{115}
\newcommand{\avhHumanAboveAgentVerbN}{18}
\newcommand{\avhSklearnVerb}{0.094}
\newcommand{\avhSklearnErosion}{0.407}
\newcommand{\avhScipyVerb}{0.145}
\newcommand{\avhScipyErosion}{0.441}
\newcommand{\avhFlaskVerb}{0.111}
\newcommand{\avhFlaskErosion}{0.209}
\newcommand{\avhFastapiVerb}{0.273}
\newcommand{\avhFastapiErosion}{0.219}
\newcommand{\avhTransformersVerb}{0.451}
\newcommand{\avhTransformersErosion}{0.473}
\newcommand{\avhTemporalCkpts}{13667}
\newcommand{\avhTemporalVerbRising}{256}
\newcommand{\avhTemporalVerbTotal}{378}
\newcommand{\avhTemporalVerbRisingPct}{68}
\newcommand{\avhTemporalVerbGrowthMedian}{36}
\newcommand{\avhTemporalErosRising}{103}
\newcommand{\avhTemporalErosTotal}{196}
\newcommand{\avhTemporalErosRisingPct}{53}
\newcommand{\avhTemporalPreN}{9705}
\newcommand{\avhTemporalPostN}{3962}
\newcommand{\avhTemporalPreVerbMedian}{0.155}
\newcommand{\avhTemporalPostVerbMedian}{0.166}
\newcommand{\avhAgentEraPreVerbMedian}{0.156}
\newcommand{\avhAgentEraPostVerbMedian}{0.168}
\newcommand{\avhAgentEraPreErosMedian}{0.278}
\newcommand{\avhAgentEraPostErosMedian}{0.319}
\newcommand{\avhAgentEraBothN}{321}
\newcommand{\avhAgentEraVerbShiftMedian}{0.002}
\newcommand{\avhAgentEraErosShiftMedian}{0.014}
\newcommand{\avhAgentEraVerbHigherPct}{55}
\newcommand{\avhAgentEraErosHigherPct}{59}
\newcommand{\avhAgentVerbSlopeMedian}{0.0144}
\newcommand{\avhAgentErosSlopeMedian}{0.0264}
\newcommand{\avhHumanVerbSlopeMedian}{0.0022}
\newcommand{\avhHumanErosSlopeMedian}{0.0053}
\newcommand{\avhAgentVerbSlopeRatio}{6.6}
\newcommand{\avhAgentErosSlopeRatio}{5.0}

\newcommand{\psCodexAntiSlopVerbPctRed}{32.8}
\newcommand{\psCodexAntiSlopEroPctRed}{34.3}
\newcommand{\psGptFiveFourAntiSlopVerbPctRed}{27.5}
\newcommand{\psGptFiveFourAntiSlopEroPctRed}{45.1}
\newcommand{\psGptFiveFiveAntiSlopVerbPctRed}{35.6}
\newcommand{\psGptFiveFiveAntiSlopEroPctRed}{57.6}
\newcommand{\psRqInitialEroReductionMax}{62.3}
\newcommand{\psRqInitialVerbReductionMax}{34.8}
\newcommand{\psRqQualityVelocityPp}{1.3}
\newcommand{\psRqCostIncreaseMean}{12.1}

\newcommand{\psLiftStrictDropAntiSlop}{2.4}
\newcommand{\psLiftStrictDropPlanFirst}{3.6}

\newcommand{\psLiftCostIncreasePct}{12.1}

\definecolor{diffstart}{named}{codebg}
\definecolor{diffincl}{named}{codegreen}
\definecolor{diffrem}{named}{red}
\lstdefinelanguage{diff}{
  basicstyle=\ttfamily\small,
  morecomment=[f][\color{diffstart}]{@@},
  morecomment=[f][\color{diffincl}]{+\ },
  morecomment=[f][\color{diffrem}]{-\ },
}
\begin{document}

\maketitle

\begin{abstract}
Software development is iterative, yet agentic coding benchmarks hide design issues through their single-shot setup.
Recent iterative benchmarks attempt to remedy this but heavily constrain an agent's design decision space, making it impossible to faithfully measure how their decisions shape future extensions.
We introduce SlopCodeBench, a benchmark of 36 problems and 196 checkpoints where agents repeatedly extend their own solutions. Unlike prior iterative benchmarks, our evolving specifications demand architectural decisions but leave internal structure to the agent.
We measure two forms of degradation: structural erosion (concentrated complexity) and verbosity (redundant code).
Evaluating 15 coding agents across open and closed models, we find that no agent fully solves any problem end-to-end, and the best agent passes 14.8\% of checkpoints.
Quality degrades across checkpoints, with structural erosion rising in 77\% of trajectories and verbosity in 75.5\%.
Compared to 473 open-source Python repositories, agent code is 2.3x more verbose and 2.0x more eroded, and the human repositories degrade less often and by smaller margins across their git histories.
Explicit quality guidance reduces initial verbosity and erosion by up to a third, without affecting degradation rates.
SlopCodeBench provides the first measurement of code degradation under iterative extension, revealing that agents pass checkpoints while producing code that erodes and bloats with each turn.
\end{abstract}

\section{Introduction}
\label{sec:introduction}

Every design decision in software engineering is a compromise with unknown future requirements. A code search program built around regular expressions works until the specification demands structural pattern matching, at which point the entire architecture must be rewritten. Existing coding-agent benchmarks systematically undermeasure this failure mode, evaluating models once against complete task specifications~\citep{swe-bench,projdevbench,vibe-code-bench,swe-rebench-v2}. They measure whether an agent can produce correct code for the current specification, not whether that code remains extensible under future change.

\begin{figure}[t]
    \centering
    \includegraphics[width=1.00\textwidth]{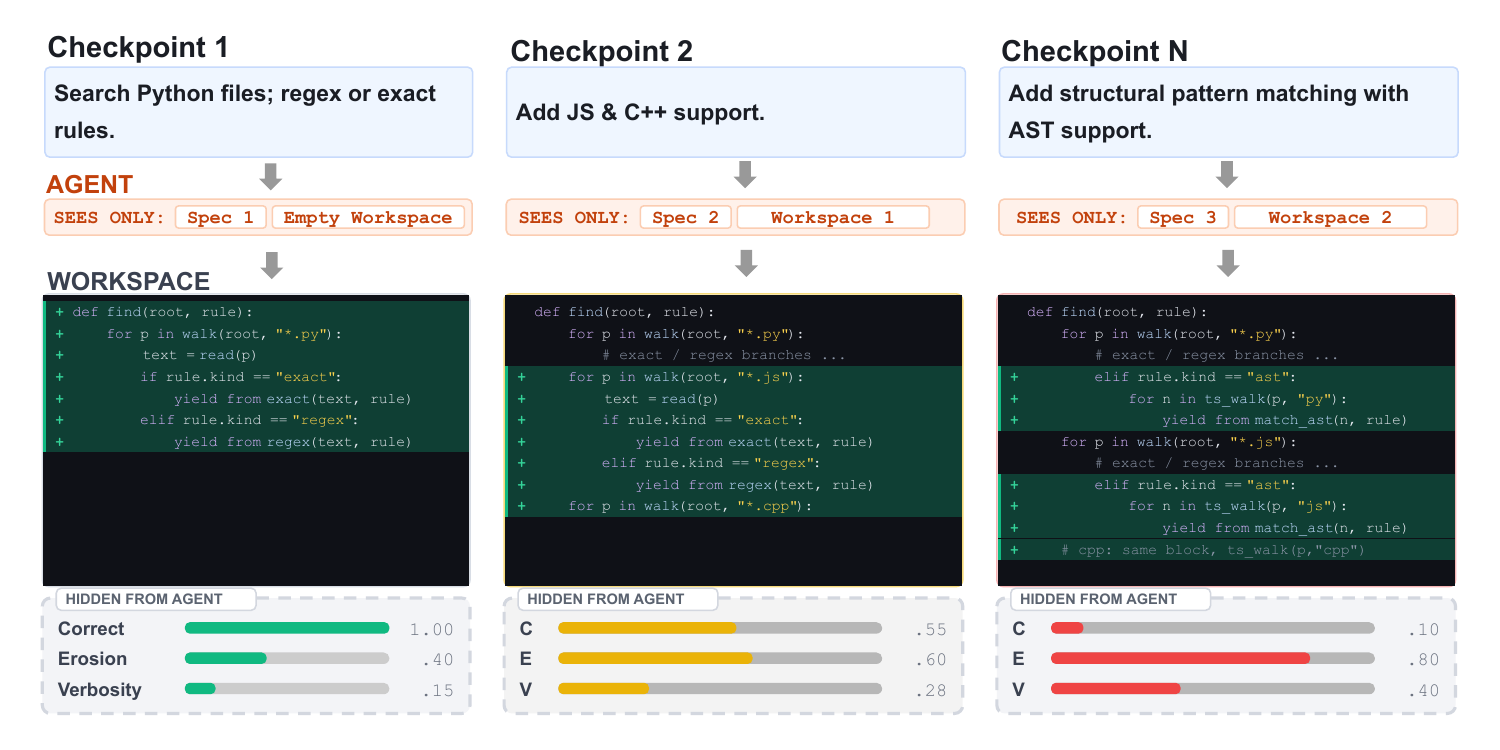}
    \caption{\textbf{Iterative evaluation in SCBench.} A task unfolds over checkpoints. 
At each step, the agent extends its prior workspace to satisfy an updated 
spec (orange). Correctness, erosion, and verbosity are scored externally 
and withheld from the agent (gray). Local shortcuts compound: correctness 
falls from $1.00$ to $0.10$ while erosion and verbosity grow.}
    \label{fig:overview}
\end{figure}

Under repeated editing, agent-generated code often deteriorates in recognizable ways. LLMs favor verbose constructions over concise idioms~\citep{dou2024-whats-wrong-llm-code, abbassi2025-taxonomy-inefficiencies}, and agent-generated code carries redundant or unnecessary methods that reviewers routinely strip~\citep{chen2025-patch-quality, nakashima2026-agentic-prs, what-to-cut}. The resulting low-quality, high-volume code is often colloquially called ``slop.'' In traditional software engineering, such accumulation is associated with higher maintenance cost and slower modification~\citep{lacerda2020-code-smells-tertiary,le2021-architectural-decay,li2022-architecture-erosion}, yet pass rates can remain stable even as the underlying code becomes harder to extend. Despite this large body of work, the standard agentic coding benchmarks cannot adequately measure this behavior because they focus on correctness in single-shot setups rather than on iterative evaluations.

Recent benchmarks push toward multi-turn or long-horizon coding, but none of them isolate \emph{true iterative coding}. Some construct iterative tasks by decomposing monolithic solutions into dependency-ordered subproblems, thereby producing a self-contained testbed rather than a realistic setting in which the agent selects an architecture and must live with it later~\citep{codeflowbench}. Others derive tasks from the commit histories of mature open-source repositories~\citep{thai2025-swe-evo,evoclaw,swe-ci}. These are valuable for studying an agent's ability to maintain or add narrow features in an existing codebase, but they fundamentally can not test iterative coding. Using human-built workspaces and historically realized evolution paths means the agent never pays the cost of its own early design decisions. In some cases, the task formulation is also tied to test- or oracle-derived signals, which further undercuts the benchmark's ability to measure open-ended iterative design~\citep{swe-ci}. To properly measure this, a benchmark needs four things: the agent builds on its own prior code; problems specify only external behavior, not internal interfaces; a truly hidden test suite to prevent leaking architectural hints; and black-box contracts that are implementable in any language.

We therefore introduce \textbf{SlopCodeBench (SCBench)}, a benchmark for measuring how code quality evolves as agents repeatedly extend \emph{their own prior code} under changing specifications. SCBench contains \transProblems{} problems spanning \overallTotalCheckpoints{} checkpoints. Each checkpoint specifies only observable behavior at a CLI or API boundary, leaving internal structure unconstrained and keeping the test suite hidden. The benchmark is language-agnostic by construction: specifications describe only CLI and API behavior. We evaluate the Python track in this paper. Beyond correctness, we track two trajectory-level quality signals: \textbf{verbosity}, which measures the growth of redundant or duplicated code, and \textbf{structural erosion}, which measures the concentration of complexity in already-complex functions. The former calculates the percent of lines that are either flagged by ast-grep patterns or are structural duplicates of other code. For erosion, we calculate the percentage of the overall cyclomatic complexity~\citep{mccabe1976-complexity} that is contained in high-complexity functions. 

Our contributions are:
\begin{enumerate}[leftmargin=*,itemsep=0pt]
    \item \textbf{\scb{}} \footnote{Full problems and code are found at \href{https://www.scbench.ai}{scbench.ai}}a language-agnostic benchmark of \transProblems{} iterative software-development problems across \overallTotalCheckpoints{} checkpoints. The state-of-the-art performance is \overallBestStrictSolveRate{}\% with no problem fully solved.
    \item \textbf{Two trajectory-level quality metrics}, verbosity and structural erosion, to measure low-quality agent code. Structural Erosion rises in \erosionPctIncreasing{}\% of trajectories and verbosity in \vfTrajIncPct{}\%.
    \item \textbf{Calibration against human code.} Agent code is \avhAllVerbRatio{}$\times$ more verbose and \avhAllErosionRatio{}$\times$ more eroded than \avhPanelN{} open-source repositories. Per checkpoint, agent verbosity grows \avhAgentVerbSlopeRatio{}$\times$ and erosion grows \avhAgentErosSlopeRatio{}$\times$ faster than that in human code.
    \item \textbf{Study on prompting techniques.} Quality-aware prompts reduce initial verbosity but do not slow the degradation. Further, this results in an average \psRqCostIncreaseMean\% increase in the cost per checkpoint and a drop in correctness of 2.3 pp.
\end{enumerate}
\section{SlopCodeBench}
\label{sec:benchmark}

\scb{} contains \transProblems{} language-agnostic problems spanning \overallTotalCheckpoints{} checkpoints. Each problem is specified solely in terms of observable behavior at a CLI or API boundary, so it can be evaluated in any implementation language. An agent implements the first specification from scratch, then repeatedly modifies and extends \emph{its own prior code} as specifications evolve. Our focus is on measuring both correctness and code quality throughout that trajectory. The remainder of this section is structured as follows: 1) our desiderata, 2) how we constructed the dataset, 3) how we quantify quality issues, and 4) finally, conclude with our evaluation procedure. 

\begin{runningexample}[Running example: \href{https://www.scbench.ai/problems/code_search}{\codesearch{}}]
\label{sec:running-example}
The agent is tasked with building a CLI tool for semantic source-file search, inspired by \href{https://ast-grep.github.io}{ast-grep}, across five checkpoints:
\begin{itemize}[leftmargin=2.2em, topsep=2pt, itemsep=1pt]
    \item[$\ckpt{1}$ -] Python files only exact and regex matching. Establishes the core CLI contract.
    \item[$\ckpt{2}$ -] Support for Javascript and C++.
    \item[$\ckpt{3}$ -] AST-based pattern matching with metavariable capture.
    \item[$\ckpt{4}$ -] Selector rules and auto-fix functionality.
    \item[$\ckpt{5}$ -] Add support for Go, Rust, and Java.
\end{itemize}

An agent that hardcodes language-specific logic at $\ckpt{1}$ faces cascading rewrites at $\ckpt{2}$ and $\ckpt{5}$; one that builds an extensible parser interface does not.
\end{runningexample}
\subsection{Design Principles}\label{sec:design-principles}

The core goal of \scb{} is to evaluate an agent's ability to make design decisions that will directly influence the quality of their code. For this, we have a core set of design principles that every problem must follow. Without these, architectural details are leaked to the agent and corrupt the signal we explicitly want to measure. They are:

\begin{enumerate}[leftmargin=*]
    \item \textbf{No prescribed internal interfaces.} Unlike existing benchmarks~\citep{swe-ci,evoclaw}, we only specify the external contract (CLI arguments or API I/O), so that the agent is forced to make \emph{architectural decisions}.
    \item \textbf{No visible test suite.} The dominant SWE evaluation paradigm provides fail-to-pass tests~\citep{swe-bench,swe-bench-plus}. \scb{} agents see only specification prose and embedded examples, never the actual test suite or its feedback. They must infer unstated edge cases from the specification alone.
    \item \textbf{Black-box, language-agnostic problem design.} Problems constrain only observable behavior, not implementation language or ecosystem. Following the principle that evaluation should not depend on a specific language's ecosystem~\citep{multi-swe-bench, swe-polybench, swe-rebench-v2}, outputs are evaluated solely via CLI or API interfaces, with normalization removing inconsequential differences in formatting and ordering. We evaluate only on Python due to cost.
\end{enumerate}

In our running \hyperref[sec:running-example]{\codesearch{}} problem, $\ckpt{1}$ specifies the CLI contract: \texttt{<root\_dir> --rules <file> [--encoding <name>]}, with output as JSON lines containing fields \texttt{rule\_id}, \texttt{file}, \texttt{start}, \texttt{end}, and \texttt{match}. The only prescribed internals are the input/output structures that the harness needs to supply inputs and parse outputs. Specifications add normalization guidance only where arbitrary choices could cause false failures, such as key ordering, text casing, or match-span sorting. In $\ckpt{3}$, for example, an example fixes the sort order for multiple pattern matches even though the rule is not stated explicitly. This prevents penalizing inconsequential implementation choices.

\subsection{Task Formulation}\label{sec:eval-setup}

All of our problems are written by hand, either as novel tasks or inspired by popular repositories. Each problem $\problem$ is an ordered list of checkpoints $[\ckptinit,\ldots,\ckpt{n}]$. At $\ckpt{i}$, the agent $\agent$ receives only the $\spec{i}$ and its previous workspace $\solution{i-1}$, then produces an updated workspace $\solution{i}$. At $\ckpt{1}$, the agent starts from the empty workspace $\solutioninit$.
\begin{equation}\label{eq:problem-setup}
    \begin{aligned}
        \solution{1} &= \agent(\spec{1}, \solutioninit) \\
        \solution{2} &= \agent(\spec{2}, \solution{1})\\
        \ldots \\
        \solution{i} &= \agent(\spec{i}, \solution{i-1})
    \end{aligned}
\end{equation}
Each checkpoint is a fresh feature starting from the prior checkpoint's workspace. The agent must reason about changes solely from the code's current structure, as we do not provide the prior conversation's context. A poor architectural choice at checkpoint $i$ becomes the foundation for checkpoint $i+1$, and the agent must build on it. If a reference solution replaces the agent's code between turns, the causal chain from early decisions to later degradation is removed. In \scb{}, unlike other benchmarks~\citep{codeflowbench, maintaincoder, evoclaw}, the agent's workspace is carried forward to the next checkpoint, and quality is measured at every step.

\paragraph{Problem Construction.} Problems were authored by the paper's authors through a two-phase process, with each problem reviewed by at least one author other than its drafter. In the proposal phase, an author drafted a problem along with its checkpoint partition; problems that did not meaningfully test design decisions, or that frontier agents could solve in a single shot, were removed from the pool. In the validation phase, we wrote the initial test suite and attempted each checkpoint with an agent, using these runs to identify and refine ambiguous or under-specified test cases. A final review pass confirmed that every problem was solvable in principle and that test suites matched the specifications. \autoref{tab:problem-design-decisions-a} details each problem and what design decisions are stressed.

\subsection{Measuring Code Quality}\label{sec:quality-dims}

Standard quality models decompose software quality into broad characteristics such as
maintainability, reliability, and portability~\citep{iso25010}. We focus on two
trajectory-level failure modes that are computable at every checkpoint and comparable
across agents and problems. \emph{Structural erosion} captures the concentration of
control-flow complexity in already-complex functions. \emph{Verbosity} captures
duplicated and rule-flagged redundant code patterns. These metrics are computed
independently of functional correctness, letting us measure structural changes that
pass rates alone do not expose.

\paragraph{Structural erosion.} Agents under iteration tend to patch new logic into existing functions rather than distributing it across focused callables. \autoref{lst:dispatch-erosion} shows how a function in \codesearch{} at $\ckpt{5}$ can grow to 117 lines with branches for each rule kind and each match source chained inside one loop. In the iterative paradigm, the clearest notion of erosion is haphazard edits to a function that patch functionality. These edits compound slowly until they become too large to work on. Thus, we define erosion as \textbf{the fraction of the codebase's total complexity mass that resides in high-complexity functions}. To this end we first assign each callable a \emph{complexity mass} that accounts for both its cyclomatic complexity (CC,~\citep{mccabe1976-complexity}) and its size:
\begin{equation}
\label{eq:mass} 
\text{mass}(f) = \text{CC}(f) \times \sqrt{\text{SLOC}(f)}
\end{equation}
where $\text{CC}(f)$ is the cyclomatic complexity of callable $f$ and $\text{SLOC}(f)$ is its source lines of code. The square root compresses the size factor so that complexity dominates rather than pure lines of code. Erosion is then the share of total mass held by functions exceeding a high-complexity threshold:
\begin{equation}
\label{eq:erosion}
\text{Erosion}
=  \frac{\sum_{f \in \mathcal{F}} \mathbb{I}[\text{CC}(f) > 10]\cdot\text{mass}(f)}{\sum_{f \in \mathcal{F}} \text{mass}(f)}
\end{equation}
where $\mathcal{F}$ is the set of all callables. We use a cutoff of 10 for CC following the established bounds in the popular code analysis tool \href{https://radon.readthedocs.io/en/latest/}{Radon}. In the solution to \codesearch{} shown in \autoref{lst:dispatch-erosion}, the majority of decision-points have collapsed into \texttt{find\_matches\_in\_file()}, driving its mass upward even as the agent adds other functions around it.

\paragraph{Verbosity.} The other dimension of slop is code that is too \emph{verbose}, copy-pasted, or unnecessary lines that do not add anything to the overall codebase. A typical \codesearch{} fragment rebuilds the array rather than using \texttt{filter}, guards each iteration with empty-list checks, and assigns intermediate names used once. To capture both effects, we use a static verbosity score with two parts. First, we measure clear patterns of wasteful code generated by agents through constructing 137 targeted \href{https://ast-grep.github.io}{AST-Grep} rules. These rules are based on observed cases of verbose code, best practices, and commonly cited anti-patterns on social media. Second, we measure structural duplication as clone lines normalized by LOC~\citep{juergens2009code,Alam2023GPTCloneBenchA}. The resulting score is
\begin{equation}
\label{eq:verbosity}
\text{Verbosity}
= \frac{|\{\text{AST-Grep Flagged Lines} \cup \text{Clone Lines}\}|}{\text{LOC}}
\end{equation}
We deduplicate lines hit by multiple AST-grep rules before counting. This score is bounded in $[0,1]$, thus comparable across runs and independent of erosion. The two metrics measure different failure modes, so tracking both gives a fuller picture of the ``slop'' agents generate.

\paragraph{Alternative Metrics.}  A natural alternative is to use existing software quality metrics directly. Composite metrics such as the Maintainability Index combine traditional measures into a single maintainability score~\citep{oman1992maintainability}. While useful as broad summaries, they are less appropriate for measuring specific ways code changes over iterations. More targeted metrics also capture only part of the phenomenon: Cyclomatic Complexity measures independent control-flow paths~\citep{mccabe1976-complexity}, Cognitive Complexity penalizes control-flow breaks and nesting~\citep{campbell2018cognitive}, the Chidamber-Kemerer suite measures class-level object-oriented design properties~\citep{chidamber1994metrics}, clone metrics capture duplication~\citep{juergens2009code,Alam2023GPTCloneBenchA}, and LOC captures code growth. In contrast, our goal is to separate two independent failure modes: code can become more verbose without concentrating complexity, and complexity can concentrate in some functions without much overall growth. We therefore measure structural erosion and verbosity separately rather than relying on a single aggregate maintainability score or fine-grained yet incomplete single-property metrics.

\subsection{Evaluating Solutions}
Every checkpoint's tests interact with the solution only through \texttt{subprocess} or its served API. Test suites normalize outputs where needed and maintain held-out tests beyond the specification's examples. Each test is categorized as:
\begin{itemize}
    \item \textbf{Core} --- Functionality explicitly mentioned or shown in the specification.
    \item \textbf{Error} --- Failure-mode behaviors.
    \item \textbf{Functionality} --- Hidden tests that exhaustively check correctness.
    \item \textbf{Regression} --- All tests from prior checkpoints. $\ckpt{1}$ has no regression tests.
\end{itemize}

The produced workspace $\solution{i}$ is \textbf{correct} if all tests pass. Because regression tests carry earlier requirements forward, a mistake at $\ckpt{2}$ can zero out later checkpoints even if later code partly works. To separate implementation quality from cascading failures, we also report \textbf{correct in isolation (ISO)} if $\solution{i}$ passes all non-regression tests for $\ckpt{i}$, and \textbf{core correct (CORE)} if it passes only the core tests. A problem is \textbf{Partially} solved if at least one checkpoint is strictly solved.

When an agent fails or crashes mid-problem, remaining checkpoints receive a correctness score of zero. Erosion and verbosity are computed only for checkpoints where the agent produced a workspace; missing checkpoints are excluded rather than imputed.

\section{Iterative Evaluation of Coding Agents}\label{sec:results}

This section reports empirical results on \scb{}. The evaluation targets
behaviors that arise once an agent edits its own code across multiple
checkpoints, a regime that one-shot benchmarks do not exercise. Four
research questions structure the experiments, each stated with its
supporting result.

\noindent\textbf{RQ1: Can agents iteratively extend their own design decisions?} No configuration is able to produce a solution that passes every checkpoint for any problem. Each checkpoint's cost rises without corresponding improvements to correctness.

\noindent\textbf{RQ2: Does iteration cause agent code to degrade?} Structural erosion increases in \erosionPctIncreasing\% of trajectories while verbosity increases in \vfTrajIncPct\% of trajectories.

\noindent\textbf{RQ3: How does agent degradation compare to repository histories?} Compared to \avhPanelN{} Python repositories, agents produce code that is \avhAllErosionRatio{}$\times$ more structurally eroded and \avhAllVerbRatio{}$\times$ more verbose. Agents additionally exhibit significantly higher degradation, especially compared to pre-2024 commits.

\noindent\textbf{RQ4: Can better prompts stop this behavior?} Better prompts can reduce initial structural erosion by up to \psRqInitialEroReductionMax\% and verbosity by up to \psRqInitialVerbReductionMax\%, but do not stop iterative degradation. The average quality velocity remains \psRqQualityVelocityPp{} percentage points per checkpoint. Different prompting methods additionally increase the cost per checkpoint by \psRqCostIncreaseMean\% on average.

\paragraph{Setup.} 

Each checkpoint runs in a fresh Docker container as a non-root user; only the working directory persists between checkpoints. Frontier models are trained for their provider's harness rather than generalized agent loops, and most developers interact with agents through these CLI tools. We therefore evaluate in native harnesses rather than frameworks such as SWE-agent~\citep{swe-agent}. We evaluate \overallNumModels{} models across six providers in detail. Each run has a two-hour wall-clock limit, no turn or cost cap, and uses the \texttt{just-solve} prompt, which only instructs the agent to solve the problem and keep track of dependencies. Additional setup details are found in \autoref{sec:eval-env-details}.

Problems range from 3 to 8 checkpoints, so raw checkpoint indices are not directly comparable across problems. For aggregation and visualization, we map each trajectory to one of five \emph{progress phases}. The first checkpoint is always \textbf{Start} and the last is always \textbf{Final}. The remaining interior checkpoints are divided into three groups: \textbf{Early}, \textbf{Mid}, and \textbf{Late}. All per-phase statistics in this paper use this.

\subsection{Solve Rates Stagnate as Cost Grows}
\label{sec:overall-results}

\begin{table}[t]
\centering
\caption{Per-model performance on \scb{}. For each model we report its best just-solve run, ranked by isolated solve rate with core, strict, and \$/checkpoint as tiebreakers. Solve rates use a fixed 196-checkpoint denominator; missing checkpoints count as unsolved. Other metrics are only for ran checkpoints. \textbf{Bold} marks the best value per column. See \autoref{tab:performance-appendix} for the full set of runs.}
\label{tab:performance-overview}
\small
\setlength{\tabcolsep}{4pt}
\resizebox{\textwidth}{!}{%
\begin{tabular}{l | rrrr | rrr | rr}
\toprule
 & \multicolumn{4}{c|}{Solve Rate (\%)}
 & \multicolumn{3}{c|}{Cost \& Time}
 & \multicolumn{2}{c}{Quality} \\
\cmidrule(lr){2-5} \cmidrule(lr){6-8} \cmidrule(l){9-10}
Model & Strict & Iso. & Core & Partial & \$/CKPT & Net \$ & Min/CKPT & Erosion & Verbosity \\
\midrule
Opus 4.5 & 9.2 & 17.3 & 57.7 & 25.0 & 2.53{\scriptsize$\pm$1.57} & 492.40 & 7.2{\scriptsize$\pm$3.7} & 0.70{\scriptsize$\pm$0.20} & 0.43{\scriptsize$\pm$0.18} \\
Opus 4.6 & 9.7 & 20.9 & \textbf{67.3} & 27.8 & 3.17{\scriptsize$\pm$2.82} & 620.80 & 13.1{\scriptsize$\pm$12.9} & 0.75{\scriptsize$\pm$0.15} & 0.44{\scriptsize$\pm$0.19} \\
Opus 4.7 & 8.2 & 20.9 & 65.8 & 36.1 & 2.17{\scriptsize$\pm$1.83} & 425.34 & 6.8{\scriptsize$\pm$5.0} & 0.76{\scriptsize$\pm$0.17} & 0.48{\scriptsize$\pm$0.21} \\
Sonnet 4.6 & 7.1 & 16.8 & 57.7 & 27.8 & 1.96{\scriptsize$\pm$1.97} & 383.83 & 14.2{\scriptsize$\pm$14.2} & 0.75{\scriptsize$\pm$0.20} & 0.44{\scriptsize$\pm$0.19} \\
\midrule
GPT~5.2 Codex & 9.7 & 21.9 & 56.1 & 36.1 & 0.85{\scriptsize$\pm$0.80} & 167.29 & 12.3{\scriptsize$\pm$9.7} & 0.73{\scriptsize$\pm$0.18} & 0.50{\scriptsize$\pm$0.16} \\
GPT~5.3 Codex & 11.2 & 26.0 & 60.7 & 41.7 & 0.66{\scriptsize$\pm$0.49} & 129.65 & 7.7{\scriptsize$\pm$4.6} & 0.64{\scriptsize$\pm$0.19} & 0.46{\scriptsize$\pm$0.18} \\
GPT~5.3 Spark & 3.1 & 8.2 & 29.1 & 11.1 & \textbf{0.20{\scriptsize$\pm$0.41}} & \textbf{37.17} & \textbf{2.4{\scriptsize$\pm$4.6}} & 0.68{\scriptsize$\pm$0.19} & 0.48{\scriptsize$\pm$0.16} \\
GPT~5.4 & 10.7 & 23.5 & 62.8 & 36.1 & 0.72{\scriptsize$\pm$0.51} & 141.32 & 8.0{\scriptsize$\pm$4.7} & 0.51{\scriptsize$\pm$0.19} & 0.33{\scriptsize$\pm$0.13} \\
GPT~5.4 Mini & 5.1 & 13.8 & 53.1 & 22.2 & 0.45{\scriptsize$\pm$0.35} & 89.03 & 6.2{\scriptsize$\pm$3.8} & 0.66{\scriptsize$\pm$0.17} & 0.42{\scriptsize$\pm$0.12} \\
GPT~5.5 & \textbf{14.8} & \textbf{28.1} & 66.8 & \textbf{50.0} & 1.51{\scriptsize$\pm$0.81} & 295.59 & 6.0{\scriptsize$\pm$2.4} & \textbf{0.49{\scriptsize$\pm$0.20}} & \textbf{0.32{\scriptsize$\pm$0.12}} \\
\midrule
Composer~2 & 6.1 & 16.3 & 52.6 & 13.9 & 0.44{\scriptsize$\pm$0.44} & 86.31 & 8.8{\scriptsize$\pm$8.3} & 0.72{\scriptsize$\pm$0.15} & 0.45{\scriptsize$\pm$0.17} \\
GLM~5.1 & 9.7 & 13.8 & 40.3 & 30.6 & 1.47{\scriptsize$\pm$1.45} & 288.85 & 31.4{\scriptsize$\pm$20.0} & 0.71{\scriptsize$\pm$0.17} & 0.41{\scriptsize$\pm$0.18} \\
Kimi~K2.5 & 4.6 & 9.7 & 39.8 & 19.4 & 0.33{\scriptsize$\pm$0.25} & 63.21 & 14.5{\scriptsize$\pm$10.0} & 0.72{\scriptsize$\pm$0.20} & 0.44{\scriptsize$\pm$0.20} \\
Kimi~K2.6 & 10.7 & 18.9 & 51.0 & 33.3 & 0.74{\scriptsize$\pm$0.64} & 133.18 & 26.9{\scriptsize$\pm$20.7} & 0.76{\scriptsize$\pm$0.20} & 0.51{\scriptsize$\pm$0.22} \\
MiniMax~M2.7 & 2.0 & 4.1 & 28.1 & 8.3 & 0.34{\scriptsize$\pm$0.23} & 65.79 & 11.9{\scriptsize$\pm$6.8} & 0.73{\scriptsize$\pm$0.16} & 0.47{\scriptsize$\pm$0.17} \\
\bottomrule
\end{tabular}%
}
\end{table}

\autoref{tab:performance-overview} summarizes solve rates across \overallNumModels{} models on \scb. No agent \emph{fully} solves any of the \transProblems{} problems. No run passes every test at every checkpoint end-to-end. \textbf{\overallBestModel{}} achieves the highest strict solve rate at \overallBestStrictSolveRate{}\% and isolated solve rate peaks at \overallIsoSolveMax{}\%. Opus 4.5 achieves the highest core solve rate at \overallCoreSolveMax{}\%.
\begin{figure}[t]
    \centering
    \includegraphics[width=\textwidth]{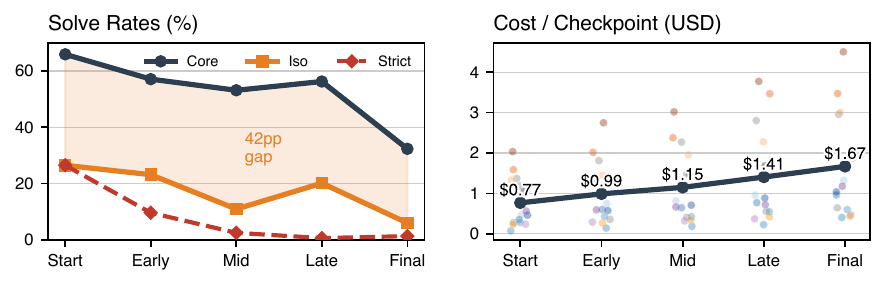}
    \caption{\textbf{Performance degrades as problems progress and cost per checkpoint steadily rise.} \emph{Left:} Checkpoint solve rate by type. \emph{Right:} Mean cost per checkpoint as problems progress.}
    \label{fig:solve-cost}
\end{figure}

\autoref{fig:solve-cost} shows the gap between core and isolated pass rates widens from \abFigCoreIsoRatioMin{}$\times$ to \abFigCoreIsoRatioMax{}$\times$. Core pass rate falls from \overallTestCorePassEarlyPct{}\% to \overallTestCorePassLatePct{}\%. Functionality pass rate changes by less than six points, from \overallTestFuncPassEarlyPct{}\% to \overallTestFuncPassLatePct{}\%, while error pass rate drops from \overallTestErrorPassEarlyPct{}\% to \overallTestErrorPassLatePct{}\%.

Cost and context grow while code movement declines. Mean cost per checkpoint grows \abFigCostGrowth{}$\times$ from the start to end, while mean relative lines changed fall from \overallChurnMeanEarlyPct{}\% early to \overallChurnMeanLatePct{}\% late. Total recorded token accounting is \overallTokensTotalB{}B tokens. Normalized to 196 checkpoints, \overallTokenLowestModel{} has the lowest token spend at \overallTokenLowestNormB{}B tokens, while \overallTokenHighestModel{} has the highest at \overallTokenHighestNormB{}B tokens. 
\subsection{Iterative Agent Trajectories Accumulate Quality Issues}\label{sec:iteraitve-degrade}

\begin{figure}[ht]
    \centering
    \makebox[\textwidth][c]{%
        \includegraphics[width=1.15\textwidth]{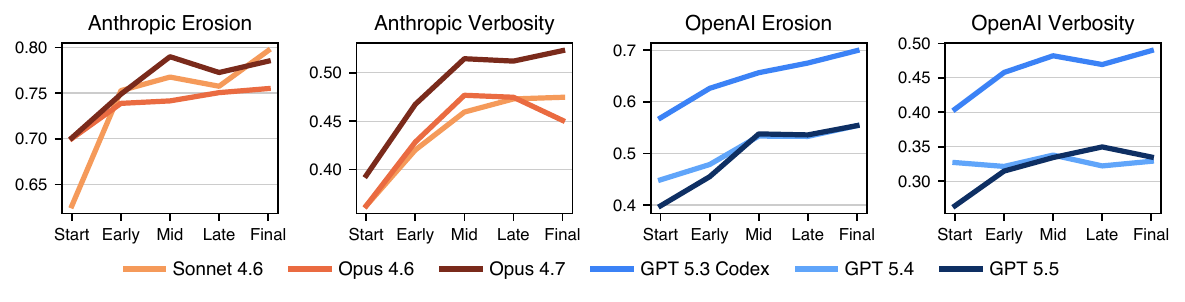}%
    }
    \caption{\textbf{Models from both frontier labs produce similar degradation trends as the problem progresses}. Erosion and verbosity across problem progress for six representative models.}
    \label{fig:accumulation-results}
\end{figure}

\autoref{fig:accumulation-results} highlights the changes in quality throughout frontier models' trajectories on problems. Across all settings, structural erosion increases in \erosionPctIncreasing\% of trajectories, verbosity in \vfTrajIncPct\%. The average number of functions with at least 10 cyclomatic complexity rises from \erosionHighCcCountFirstMean{} to \erosionHighCcCountLastMean{} and mean maximum CC from \newErosionFirstCcMaxMean{} to \newErosionLastCcMaxMean{}. This behavior would be invisible without iterative evaluation.

Opus~4.7's \texttt{main()} on \href{https://www.scbench.ai/problems/forge}{\texttt{forge}} grows \erosionForgeMainCcMultiplier$\times$ in CC across \erosionForgeNumCkpts{} checkpoints, from \erosionForgeMainLinesFirst{} to \erosionForgeMainLinesLast{} lines. The $\ckpt{1}$ \texttt{argparse} call, shown in \autoref{lst:erosion-main-cp1}, is replaced by a manual flag loop in which seven flags each get an identical pair of \verb|--X val|/\verb|--X=val| branches and no helper is extracted (\autoref{lst:erosion-main-cp8}). The accumulation over time highlights the key failure mode.

Verbosity behaves similarly to erosion. Structural duplication grows \vfCloneGrowthPct\% while AST-grep violation density grows only \vfAstOverallGrowth\%, reflecting the rewriting of existing code rather than fresh rule violations. \textbf{Iterative evaluation makes accumulation visible; small structural choices are amplified by iteration rather than absorbed.}

\subsection{How does agent degradation compare to repository histories?}
\label{sec:agent-vs-human}

\begin{figure}[t]
    \centering
    \includegraphics[width=\linewidth]{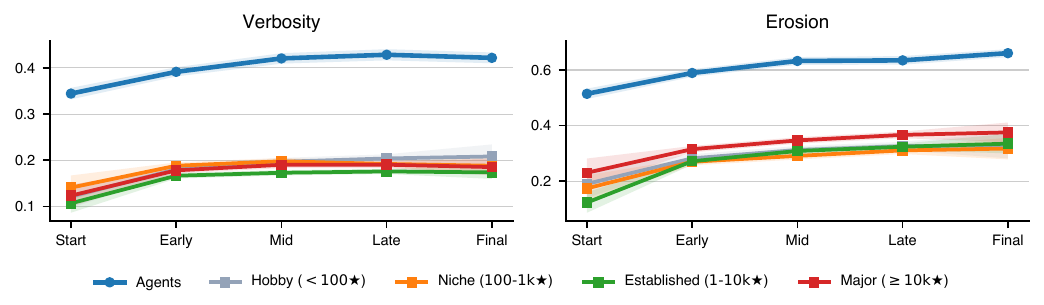}
    \caption{\textbf{Agents produce lower quality code and degrade it faster than real developers.} Verbosity and erosion across normalized trajectory progress, with 95\% confidence intervals. GitHub-star bins over \avhPanelN{} repositories for a total of $\avhTemporalCkpts$ checkpoints. \autoref{tab:agent-vs-human} has HEAD commit details.}
    \label{fig:agent-human-trajectory}
\end{figure}

To answer this, we collect \avhPanelN{} Python repositories ranging from hobby projects under 100 stars to major frameworks with over 10k stars and 2M LOC, covering web frameworks, scientific computing, infrastructure, and utilities. For each repository, we randomly sample up to 30 source-modifying commits totaling \avhTemporalCkpts{} commits. We treat each repository's most recent commit at the time of collection as its current state for cross-sectional comparisons and the full commit sequence as analogous to an agent trajectory.

\autoref{fig:agent-human-trajectory} shows the human panel averages 0.19 verbosity and 0.34 erosion. Agent checkpoints sit \avhAllVerbRatio{}$\times$ and \avhAllErosionRatio{}$\times$ above those means. Only \avhHumanAboveAgentVerbN{} of \avhPanelN{} repositories cross the agent verbosity mean. Agent verbosity grows by \avhAgentVerbSlopeMedian{} per checkpoint and erosion by \avhAgentErosSlopeMedian{}; the human medians are \avhHumanVerbSlopeMedian{} and \avhHumanErosSlopeMedian{}. Inside the iteration loop, agents accumulate verbosity roughly 7$\times$ faster and erosion 5$\times$ faster. Erosion rises in \erosionPctIncreasing{}\% of agent trajectories against \avhTemporalErosRisingPct{}\% of human repositories.

\begin{table}[t]
    \centering
    \caption{Verbosity and erosion at HEAD commit, by GitHub-star tier and against agent checkpoints.}
    \label{tab:agent-vs-human}
    \footnotesize
    \setlength{\tabcolsep}{6pt}
    \begin{tabular}{lccc}
        \toprule
        Group & $n$ & Verbosity & Erosion \\
        \midrule
        Hobby ($< 100\bigstar$) & 114 & $0.21 \pm 0.14$ & $0.33 \pm 0.25$ \\
        Niche ($100\text{--}1\text{k}\bigstar$) & 124 & $0.19 \pm 0.11$ & $0.32 \pm 0.22$ \\
        Established ($1\text{k}\text{--}10\text{k}\bigstar$) & 120 & $0.18 \pm 0.08$ & $0.33 \pm 0.20$ \\
        Major ($\geq 10\text{k}\bigstar$) & 115 & $0.19 \pm 0.10$ & $0.37 \pm 0.19$ \\
        \midrule
        Human panel (all) & 473 & $0.19 \pm 0.11$ & $0.34 \pm 0.22$ \\
        Agent checkpoints & 2869 & $0.44 \pm 0.18$ & $0.68 \pm 0.20$ \\
        \bottomrule
    \end{tabular}
\end{table}

For the \avhAgentEraBothN{} repositories with at least three commits in each of the pre- and post-January-2024 windows, within-repository medians shift by $+\avhAgentEraVerbShiftMedian$ for verbosity and $+\avhAgentEraErosShiftMedian$ for erosion; \avhAgentEraErosHigherPct{}\% of these repositories are more eroded after 2024 than before. The shift is consistent with a small uptick in human-authored slop after agent assistance became common, and it remains far below what an agent accumulates in a single checkpoint. \autoref{sec:human-repo-panel} extends this analysis. 
    
\subsection{Explicit Prompting Improves Initial Quality, But Still Degrades}
\label{sec:prompt-strategy}
We next evaluate whether prompt-side modifications can mitigate the degradation induced by iteration. Two prompts are tested against the \justsolve{}~ baseline. \antislop{} (\autoref{lst:anti-slop-prompt}) enumerates verbosity and over-engineering patterns the agent should avoid. \planfirst{} (\autoref{lst:plan-first-prompt}) requires the agent to produce a written plan before generating code.

\begin{figure}[t]
    \centering
    \includegraphics[width=\textwidth]{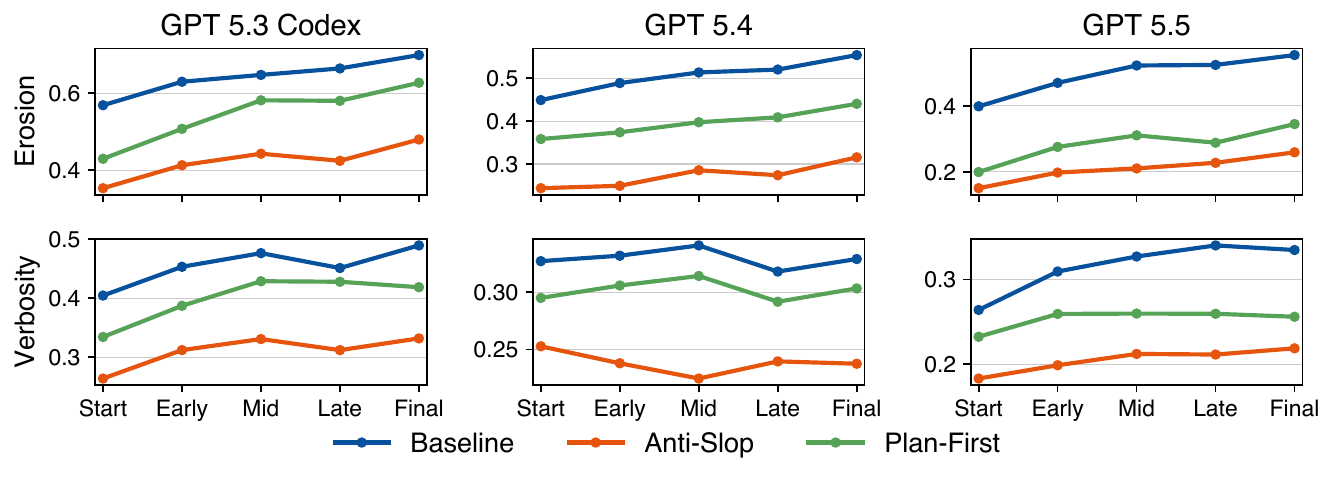}
    \caption{\textbf{Prompting can improve initial quality, but does not remove the iterative issues.} Mean structural erosion (top) and verbosity (bottom) at each progress bin, per model.}
    \label{fig:prompt-strategy-trajectories}
\end{figure}

\definecolor{good}{HTML}{1B7837}
\definecolor{bad}{HTML}{B2182B}
\begin{table}[t]
    \centering
    \caption{Raw values per prompt with the lift over \justsolve{} in parentheses. \textcolor{good}{Green} means improvements while \textcolor{bad}{red} indicates regressions.}
    \label{tab:prompt-strategy-lift}
    \footnotesize
    \setlength{\tabcolsep}{5pt}
    \begin{tabular}{llcccccc}
        \toprule
        Model & Prompt & Strict & Iso. & Core & Erosion & Verbosity & Cost \\
        \midrule
        \multirow{3}{*}{GPT 5.3 Codex} & Just-Solve & 11.2 & 26.0 & 60.7 & 0.64 & 0.46 & 0.66 \\
         & Anti-Slop & 11.2 (+0.0) & 23.5 (\textcolor{bad}{-2.6}) & 57.1 (\textcolor{bad}{-3.6}) & 0.42 (\textcolor{good}{-0.22}) & 0.31 (\textcolor{good}{-0.15}) & 0.73 (\textcolor{bad}{+0.07}) \\
         & Plan-First & 8.2 (\textcolor{bad}{-3.1}) & 21.4 (\textcolor{bad}{-4.6}) & 62.2 (\textcolor{good}{+1.5}) & 0.55 (\textcolor{good}{-0.10}) & 0.40 (\textcolor{good}{-0.06}) & 0.68 (\textcolor{bad}{+0.02}) \\
        \midrule
        \multirow{3}{*}{GPT 5.4} & Just-Solve & 10.7 & 23.5 & 62.8 & 0.51 & 0.33 & 0.72 \\
         & Anti-Slop & 9.2 (\textcolor{bad}{-1.5}) & 25.5 (\textcolor{good}{+2.0}) & 61.2 (\textcolor{bad}{-1.5}) & 0.28 (\textcolor{good}{-0.23}) & 0.24 (\textcolor{good}{-0.09}) & 0.90 (\textcolor{bad}{+0.18}) \\
         & Plan-First & 9.7 (\textcolor{bad}{-1.0}) & 25.0 (\textcolor{good}{+1.5}) & 63.3 (\textcolor{good}{+0.5}) & 0.39 (\textcolor{good}{-0.11}) & 0.30 (\textcolor{good}{-0.02}) & 0.83 (\textcolor{bad}{+0.11}) \\
        \midrule
        \multirow{3}{*}{GPT 5.5} & Just-Solve & 14.8 & 28.1 & 66.8 & 0.49 & 0.32 & 1.51 \\
         & Anti-Slop & 9.2 (\textcolor{bad}{-5.6}) & 24.0 (\textcolor{bad}{-4.1}) & 63.3 (\textcolor{bad}{-3.6}) & 0.21 (\textcolor{good}{-0.28}) & 0.21 (\textcolor{good}{-0.11}) & 1.68 (\textcolor{bad}{+0.17}) \\
         & Plan-First & 8.2 (\textcolor{bad}{-6.6}) & 24.0 (\textcolor{bad}{-4.1}) & 67.3 (\textcolor{good}{+0.5}) & 0.29 (\textcolor{good}{-0.21}) & 0.25 (\textcolor{good}{-0.06}) & 1.61 (\textcolor{bad}{+0.11}) \\
        \bottomrule
    \end{tabular}
\end{table}

\autoref{fig:prompt-strategy-trajectories} highlights that the magnitudes of structural erosion and verbosity are significantly lowered by prompt strategies. Under \antislop{}, verbosity is lower than baseline by \psGptFiveFourAntiSlopVerbPctRed\% on GPT~5.4, \psGptFiveFiveAntiSlopVerbPctRed\% on GPT~5.5, and \psCodexAntiSlopVerbPctRed\% on GPT~5.3~Codex. Erosion is lower by \psGptFiveFourAntiSlopEroPctRed\%, \psGptFiveFiveAntiSlopEroPctRed\%, and \psCodexAntiSlopEroPctRed\% respectively. Only with GPT 5.4 do we observe that quality \emph{improves} from start to end with the \antislop{} prompt.

This improved quality comes with some trade-offs, as highlighted in \autoref{tab:prompt-strategy-lift}. Across models, the base \justsolve~prompt has the best strict performance with an average drop of \psLiftStrictDropAntiSlop~pp for \antislop~and \psLiftStrictDropPlanFirst~pp for \planfirst. Core pass rate improves by 0.8~pp for \planfirst. Both prompts raise the cost per checkpoint by \psLiftCostIncreasePct\% on average. \textbf{Overall, prompting strategies trade capabilities for better initial quality, with little impact on the iterative degradation.}

\section{Related Work}
\label{sec:related_work}

\paragraph{Single-shot benchmarks.} Repository-level evaluation against a fixed specification began with \citet{swe-bench} and has broadened across language, domain, and task complexity~\citep{multi-swe-bench, swe-polybench, longcli-bench, fea-bench, commit0}. Pass rate hides defect categories that tests do not exercise~\citep{swe-abs, lessleak-bench}.

\paragraph{Iterative and evolutionary benchmarks.} Prior iterative work severs accumulation through per-step scoring, clean prior-turn states, isolated modifications, or unchained long-horizon framing~\citep{sr-eval, recode-h, codeflowbench, maintaincoder, humanevo, thai2025-swe-evo}; \citet{swe-ci} preserves history but draws from public repositories and inherits contamination risk. \citet{evoclaw} chains agent code across 98 milestones with pass rates falling from 80\% to 38\%, yet evaluation stays pass/fail, and SLUMP~\citep{slump} measures only final-artifact compliance. Verbosity and erosion require the agent's own output to accumulate across many steps.

\paragraph{Quality degradation and code smells.} LLM code carries higher per-line cyclomatic complexity than human code, with redundant steps, duplication, and unnecessary conditionals its most prevalent smells~\citep{dou2024-whats-wrong-llm-code, abbassi2025-taxonomy-inefficiencies, cotroneo2025-human-vs-ai}. Under repeated modification, outputs drift toward structural attractors, quality diverges across trajectories, and refinement introduces defects that tests miss~\citep{iterative-readability, chen2025-patch-quality, santos2025-software-aging, scaffold-cegis, diff-fuzzing-refactoring, bohr2025-show-and-tell, interaction-smells, overcorrection-reviewers, more-rounds-more-noise, nakashima2026-agentic-prs, vibe-coding-overhead}. Our metrics quantify these dynamics per checkpoint.

\section{Conclusion}
\label{sec:conclusion}

We present \scb{}, an iterative benchmark designed to measure the quality of code produced by agents when they are forced to make design decisions. We follow explicit construction principles to create black-box problems with only external contracts described. To measure quality, we introduce structural erosion, complexity concentration, and verbosity, the fraction of code that is redundant. No agent is able to solve any of our \transProblems{} problems end-to-end, while the state-of-the-art solve rate is only \overallBestStrictSolveRate{}\% of \overallTotalCheckpoints{} checkpoints. When forced to iterate, agents consistently degrade the codebase's quality. In our maintained-repository calibration panel, agent checkpoints exhibit higher erosion and verbosity, and agent trajectories accumulate both metrics faster than those of sampled Python repository histories. Our analysis of prompting interventions finds that they can improve initial quality issues, but cannot stop the eroding behavior. \scb{} is a first-of-its-kind iterative evaluation that allows proper investigation of how agents act when given decision autonomy.

\begin{ack}
We thank Abtin Molavi, Amanda Xu, June Cho, Xavier Garcia, and Samuel Guo for feedback during the development of this project. We also thank the Terminal Bench team for their inspiration and feedback.

This work is supported by Snorkel AI through the Open Benchmarks Grants program. This work was additionally supported in part by DARPA and the NSF.
\end{ack}

\bibliographystyle{plainnat}
\bibliography{references}
\appendix

\needspace{4\baselineskip}
\section{Benchmark Details}\label{sec:appendix-benchmark}

\subsection{Code Search Checkpoint Specifications}\label{sec:appendix-code-search}

This subsection provides the complete specifications for the first three checkpoints of the \texttt{code\_search} problem referenced in \autoref{sec:running-example}. Each specification appears exactly as presented to agents during evaluation.

\subsubsection{Checkpoint 1: Python-only Exact and Regex Matching}
\label{sec:appendix-cp1}

\begin{lstlisting}[style=specblock, caption={Specification for \texttt{code\_search} checkpoint 1}, label={lst:code-search-cp1}]
Build a command-line code searcher for Python codebases.
It takes a directory of source files and a set of rules,
then prints one JSON object per match (JSON Lines) to STDOUT.

Supported language (by file extension only): Python (.py).
Supported rule types (exactly these two): exact match and
vanilla regex.

# Deliverable

Write an executable that accepts:

  <entry_command> <root_dir> --rules <rules_file> [--encoding <name>]

* <root_dir>: path to the codebase to scan.
* --rules <rules_file>: path to a JSON array of rules (see schema).
* --encoding <name>: optional; default utf-8. Files that fail to
  decode are skipped.

Output: JSON Lines to STDOUT, one object per match (schema below).
On success (even if zero matches): exit code 0.

## Inputs

### File type -> language

Only files with extension .py are scanned. All other files are
ignored.

### Rules file (JSON array)

Each rule is an object with this schema:

{
  "id": "<non-empty string>",
  "kind": "exact" | "regex",
  "pattern": "<non-empty string>",
  "languages": ["python"],                 // optional; default: ["python"]
  "regex_flags": ["i", "m", "s"]           // optional; only for kind = "regex"
}

Constraints on types/inputs

* id: unique across the rules array.
* pattern: valid UTF-8 string. For kind="regex", it must compile
  with provided regex_flags (subset of i case-insensitive, m multiline,
  s dotall). No other flags allowed.
* languages (when present) must be an array of strings and may only
  contain "python".
* Missing optional fields use their defaults.

### Matching semantics

Matches inside comments/strings are allowed.

## Output (JSON Lines to STDOUT)

For each match, print a single JSON object with exactly:

{
  "rule_id": "<string>",
  "file": "<posix path>",                 // path relative to root_dir, '/' separators
  "language": "python",
  "start": {"line": <int>, "col": <int>}, // 1-based line/column.
  "end": {"line": <int>, "col": <int>},   // position immediately AFTER the match
  "match": "<string>"
}

No extra fields. Each object on its own line with a trailing newline.
No other STDOUT output.

## Normalization

1. Ordering: Matches must appear by file (lexicographically), then
   start.line, then start.col, then rule_id.
2. Coordinates: Lines and columns are 1-based.
3. Path format: file is relative to <root_dir> with '/' separators.
4. Encoding: Read files using --encoding (default utf-8); skip files
   that fail to decode.
\end{lstlisting}

\subsubsection{Checkpoint 2: Multi-language Support with Filtering}
\label{sec:appendix-cp2}

\begin{lstlisting}[style=specblock, caption={Specification for \texttt{code\_search} checkpoint 2}, label={lst:code-search-cp2}]
Extend your code searcher to support JavaScript and C++
source files.

---

## New Requirements

### File type -> language

Scan these extensions:

| Language   | Extensions                                   |
| ---------- | -------------------------------------------- |
| Python     | .py                                          |
| JavaScript | .js, .mjs, .cjs                              |
| C++        | .cc, .cpp, .cxx, .hh, .hpp, .hxx             |

### Rule language filtering

Rules may specify "languages" from this set:
["python", "javascript", "cpp"].
If omitted, the rule applies to all three.

---

## Example

### rules.json

[
  {"id":"todo","kind":"exact","pattern":"TODO:"},
  {"id":"printf","kind":"regex","pattern":"\\bprintf\\s*\\(","languages":["cpp"]},
  {"id":"console-log","kind":"regex","pattern":"console\\.log\\s*\\(","languages":["javascript"]}
]

### Project tree

repo/
  main.py
  app.js
  src/
    engine.cpp

### Run

$ <entry_command> repo --rules rules.json

### Output

{"rule_id":"console-log","file":"app.js","language":"javascript","start":{"line":5,"col":3},"end":{"line":5,"col":15},"match":"console.log("}
{"rule_id":"printf","file":"src/engine.cpp","language":"cpp","start":{"line":42,"col":3},"end":{"line":42,"col":10},"match":"printf("}
{"rule_id":"todo","file":"main.py","language":"python","start":{"line":1,"col":1},"end":{"line":1,"col":6},"match":"TODO:"}
\end{lstlisting}

\subsubsection{Checkpoint 3: Structure-Aware Metavariable Patterns}
\label{sec:appendix-cp3}

\begin{lstlisting}[style=specblock, caption={Specification for \texttt{code\_search} checkpoint 3}, label={lst:code-search-cp3}]
Extend your code searcher to support structure-aware patterns
with metavariables.

Supported rule kinds: exact, regex, and pattern.

# Deliverable

Your existing executable is extended to understand kind: "pattern"
in the rules file:

  <entry_command> <root_dir> --rules <rules_file> [--encoding <name>]

Output: JSON Lines (one object per match).

## Pattern Rules

### Rule schema additions

{
  "id": "<non-empty string>",
  "kind": "pattern",
  "pattern": "<code-like string with metavariables>",
  "languages": ["cpp" | "javascript" | "python"],      // optional; default: all 3
}

### Metavariables

* Any token of the form $NAME (e.g., $X, $GREETING) is a metavariable.
  * Metavariables ending in ? (e.g., $X?) are optional -- the pattern
    matches even if they are not present.
* A metavariable matches a single code element appropriate for that
  position.
* If the same metavariable name appears multiple times in the pattern,
  all occurrences must match the same text.
* $$ in the pattern matches a literal $ in the source code.

### Pattern string

* The pattern must be valid code in the target language (with
  metavariables treated as valid placeholders).
* No wildcards beyond metavariables are required (no ellipsis or
  quantifiers).

### Matching semantics

Find all matches in source files where the rule's language applies.

## Output (JSON Lines to STDOUT)

For kind: "pattern", the JSON object per match is:

{
  "rule_id": "<string>",
  "file": "<posix path>",
  "language": "cpp" | "javascript" | "python",
  "start": {"line": <int>, "col": <int>},
  "end": {"line": <int>, "col": <int>},
  "match": "<string>",
  "captures": {
    "$NAME": {
      "text": "<matched source text>",
      "ranges": [
        {"start": {"line": <int>, "col": <int>}, "end": {"line": <int>, "col": <int>}}
      ]
    }
    // ... one entry per metavariable bound in this match
  }
}

* ranges lists every occurrence of that metavariable within the matched
  region (use the same 1-based, Unicode column coordinates).
* If a metavariable appears once, ranges has a single range.

## Normalization

1. Pattern determinism: When multiple matches share the same start
   position, sort by end position (earlier end first), then by rule_id.
2. Captures key order: Serialize captures with keys sorted
   lexicographically by metavariable name (e.g., $A before $X).
\end{lstlisting}
\newpage

\begin{table*}[t]
\centering
\caption{SlopCodeBench problems and the design decisions stressed by iterative checkpoints, part 1 of 2. \textbf{CPs} gives the number of checkpoints.}
\label{tab:problem-design-decisions-a}
\scriptsize
\setlength{\tabcolsep}{4pt}
\renewcommand{\arraystretch}{1.08}
\begin{tabularx}{\textwidth}{@{}l c >{\raggedright\arraybackslash}X@{}}
\toprule
Problem & CPs & Design decisions stressed \\
\midrule
\texttt{cfgpipe} & 6 & Source-priority resolution, typed parsing boundaries, nested schema representation, watch and event architecture, redaction, and store-prefix composition. \\
\texttt{circuit\_eval} & 8 & Circuit IR, parser separation across formats, two-valued versus three-valued logic, optimization pass pipeline, and equivalence checking. \\
\texttt{code\_search} & 5 & Regex versus AST matching, metavariable representation, rewrite conflict handling, selector semantics, and language adapter boundaries. \\
\texttt{dag\_execution} & 3 & DAG DSL parsing, task parameter model, execution ordering, cache key design, and per-task cache overrides. \\
\texttt{database\_migration} & 5 & Migration operation model, rollback representation, dependency sorting, cycle errors, and constraint or index handling. \\
\texttt{datagate} & 7 & Dataset ingestion abstraction, query and filter pipeline, pagination and export contracts, cache invalidation, runtime config, and access control. \\
\texttt{dynamic\_buffer} & 4 & Example-to-transform inference, generated API shape, stateful transform classes, streaming buffers, and cache behavior. \\
\texttt{dynamic\_config\_service\_api} & 4 & Versioned config storage, inheritance and deep merge rules, schema validation, approval workflow state, and policy enforcement. \\
\texttt{env\_manager} & 5 & Declarative provisioning IR, OS-specific planning boundaries, deterministic output, module validation, profiles, and build-script generation. \\
\texttt{etl\_pipeline} & 5 & Pipeline IR, expression language design, branching, reusable sub-pipeline parameters, and library loading. \\
\texttt{eve\_industry} & 6 & SDE normalization, recipe graph representation, invention probability, ME and TE accounting, job scheduling, and recursive build planning. \\
\texttt{eve\_jump\_planner} & 3 & Route graph modeling, fuel and fatigue costs, spatial distance calculations, cloak mechanics, and high-sec fallback routing. \\
\texttt{eve\_market\_tools} & 4 & Market order ingestion, price-book aggregation, reprocessing yield math, hub optimization, and arbitrage or hauling profit ranking. \\
\texttt{eve\_route\_planner} & 3 & Warp physics, route cost calculation, cargo manifest planning, multi-trip scheduling, and contract selection under constraints. \\
\texttt{execution\_server} & 6 & Command execution boundaries, file tracking, output format abstraction, chain and hook semantics, cache keys, persistent environment concurrency, and job queue DAGs. \\
\texttt{file\_backup} & 4 & Schedule schema, backup strategy abstraction, glob exclusion semantics, verification, incremental hashing, and destination layout. \\
\texttt{file\_merger} & 4 & Multi-format reader abstraction, schema alignment, external sort strategy, partition and shard layout, and nested type flattening. \\
\texttt{file\_query\_tool} & 5 & SQL parsing and execution planning, file table abstraction, joins, aggregation, window state, CTEs, and subquery scope. \\
\bottomrule
\end{tabularx}
\end{table*}

\begin{table*}[t]
\centering
\caption{SlopCodeBench problems and the design decisions stressed by iterative checkpoints, part 2 of 2.}
\label{tab:problem-design-decisions-b}
\scriptsize
\setlength{\tabcolsep}{4pt}
\renewcommand{\arraystretch}{1.08}
\begin{tabularx}{\textwidth}{@{}l c >{\raggedright\arraybackslash}X@{}}
\toprule
Problem & CPs & Design decisions stressed \\
\midrule
\texttt{forge} & 8 & Resource models for blueprints, allocations, units, modules, revision gates, strict JSON contracts, and admin access boundaries. \\
\texttt{l2m} & 5 & LaTeX parsing depth, macro expansion, environment conversion rules, CLI flags, and KaTeX-compatible Markdown output. \\
\texttt{layered\_config\_synthesizer} & 4 & Layer merge semantics, conflict resolution, fragment expansion, environment interpolation, multi-run manifests, and JSON Schema validation. \\
\texttt{log\_query} & 5 & SQL-like grammar design, NDJSON streaming execution, aggregation and joins, schema label mapping, and subquery scope. \\
\texttt{meshctl} & 8 & Declarative resource specs, persisted state shape, lifecycle and topology operations, security model, migrations, network exposure, and multi-region policy. \\
\texttt{metric\_transform\_lang} & 5 & DSL parser and interpreter design, stream processing, aggregation and window state, temporal joins, and resumable execution. \\
\texttt{migrate\_configs} & 5 & Rule engine design, multi-format parsing, inheritance and cycle detection, array pattern matching, relocation rules, and validation order. \\
\texttt{mocked\_http} & 8 & Mock schema design, request matching, JSONPath evaluation, environment interpolation, validation strategy, and admin runtime behavior. \\
\texttt{mvvault} & 6 & Versioned catalog schema handling, field-level history, sync and download boundaries, viewer routing, annotations, and auto-migration triggers. \\
\texttt{pwd\_manager} & 5 & Vault storage and encryption boundary, unlock lifecycle, search and edit model, categories, clipboard integration, import or export, and locking behavior. \\
\texttt{recli} & 8 & CLI command tree architecture, config inheritance, aliases, output formatting, file cache and SQLite persistence, container orchestration, upgrades, and system checks. \\
\texttt{rejector} & 5 & Rate-limit scheduler, task and generation scheme abstractions, ICL setup representation, agentic tool-loop state, and multi-provider routing. \\
\texttt{sheeteval} & 7 & Rule schema for grading, formula and dependency evaluation, fatal and concealed controls, threshold semantics, scenario check mode, and HTML reports. \\
\texttt{sith} & 6 & Static versus interpreter-assisted analysis, symbol indexing, refactor safety, project root discovery, and environment or settings handling. \\
\texttt{test\_translator} & 8 & Structured test spec parsing, generator per target language, test discovery, equality semantics, generate-before-test enforcement, and loop-as-test translation. \\
\texttt{textdrop} & 6 & HTTP boundary validation, Markdown rendering, metadata and TOC extraction, static assets, auth and drain lifecycle, and storage backend abstraction. \\
\texttt{trajectory\_api} & 5 & Trajectory data model, token and cost tracking, mutable updates with ETags, fork lineage, grammar parsing for tool calls, and sandboxed execution. \\
\texttt{xjq} & 5 & Multi-format input detection, XPath and CSS selector dispatch, text extraction modes, file precedence, and XML or JSON output formatting. \\
\midrule
\multicolumn{1}{@{}l}{\textbf{Total}} & \textbf{196} & \textbf{36 problems} \\
\bottomrule
\end{tabularx}
\end{table*}

\needspace{4\baselineskip}
\section{Quality and Erosion Analysis}\label{sec:appendix-quality}

\subsection{Code Quality Examples}\label{sec:appendix-quality-examples}

This subsection reproduces the two \codesearch{} code fragments referenced in \autoref{sec:quality-dims}. \autoref{lst:dispatch-erosion} illustrates structural erosion, and \autoref{fig:verbosity-example} illustrates verbosity.

\begin{figure}[h]
\begin{lstlisting}[caption={Structurally eroded \texttt{find\_matches\_in\_file()} in \codesearch{} at $\ckpt{5}$ (117 lines total). Nearly all decision-point mass ends up in one function.},label={lst:dispatch-erosion}]
def find_matches_in_file(text, path, language, rules):
    ...
    for rule in rules:
        if kind == "exact":
            ...
        elif kind == "regex":
            ...
        elif kind == "pattern":
        else: ...

        if iterable is not None:
            for match in iterable:
                ...
        if kind == "pattern":
            for match in iter_pattern_matches(...):
                ...
        for node in iter_tree_nodes(source_root):
            if not node_matches_selector(selector, node):
                ...
    return matches
\end{lstlisting}
\end{figure}

\begin{figure}[h]
\begin{lstlisting}[caption={Overly verbose code from \codesearch{}. Identity list comprehension instead of \texttt{filter}, empty checks instead of building around iteration, and single-use variables.},label={fig:verbosity-example}]
for posix_path, full_path, language in source_files:
    applicable_rules = [
        r for r in all_compiled_rules
        if language in r["languages"]
    ]
    if not applicable_rules:
        continue
    with open(full_path, "r", encoding=encoding) as f:
        content = f.read()
    matches = find_matches_in_content(
        content, applicable_rules, language
    )
    if not matches:
        continue
    match_list.extend(matches)
all_matches = deduplicate_matches(match_list)
return all_matches
\end{lstlisting}
\end{figure}

\subsection{Erosion Example: \texttt{forge/main} Across Eight Checkpoints}\label{sec:appendix-erosion-example}

On \texttt{forge}, Opus~4.7's \texttt{main()} grows \erosionForgeMainCcMultiplier$\times$ in cyclomatic complexity over \erosionForgeNumCkpts{} checkpoints, from CC \erosionForgeMainCcFirst{} to CC \erosionForgeMainCcLast{}, and from \erosionForgeMainLinesFirst{} to \erosionForgeMainLinesLast{} lines. The increase is monotonic across every checkpoint. \autoref{lst:erosion-main-cp1} shows the $\ckpt{1}$ implementation, a six-line \texttt{argparse} setup followed by a small dispatch ladder. \autoref{lst:erosion-main-cp8} shows the $\ckpt{8}$ implementation, in which \texttt{argparse} has been removed and replaced with a manual \texttt{while}-loop that handles each flag in two parallel branches, one for the \verb|--X val| form and one for \verb|--X=val|. This pair of branches is repeated for seven flags before control passes to a nested resource-by-command dispatch ladder. No shared helper is introduced across the seven flag pairs, despite the per-flag template being uniform.

\begin{figure}[h]
\begin{lstlisting}[caption={\texttt{main()} from Opus~4.7's \texttt{forge} solution at $\ckpt{1}$. Argument parsing is delegated to \texttt{argparse}; dispatch is a single \texttt{if/elif} ladder.},label={lst:erosion-main-cp1}]
def main():
    parser = argparse.ArgumentParser(add_help=False)
    parser.add_argument("--data-dir", required=True)
    parser.add_argument("resource")
    parser.add_argument("command")
    parser.add_argument("rest", nargs=argparse.REMAINDER)
    args = parser.parse_args()

    if args.resource != "blueprint":
        emit_error("validation_error", f"unknown resource: {args.resource}")

    cmd, rest = args.command, args.rest
    if cmd == "create":   cmd_create(args.data_dir)
    elif cmd == "list":   cmd_list(args.data_dir)
    elif cmd == "get":    cmd_get(args.data_dir, rest[0])
    # ... two more branches ...
    else:
        emit_error("validation_error", f"unknown command: {cmd}")
\end{lstlisting}
\end{figure}

\begin{figure}[h]
\begin{lstlisting}[caption={\texttt{main()} from Opus~4.7's \texttt{forge} solution at $\ckpt{8}$, abridged to expose structure. The \texttt{argparse} call has been removed; each of seven flags is handled by an identical pair of branches. After parsing, a nested \texttt{if resource == ...: if cmd == ...} ladder dispatches over seven resources.},label={lst:erosion-main-cp8}]
def main():
    argv = sys.argv[1:]
    data_dir = revision_arg = config_path = role = scope = None
    rules_file_flag = None
    cleaned, i = [], 0
    while i < len(argv):
        tok = argv[i]
        if tok == "--data-dir":
            if i + 1 >= len(argv):
                emit_error("validation_error", "--data-dir requires a value")
            data_dir = argv[i + 1]; i += 2
        elif tok.startswith("--data-dir="):
            data_dir = tok[len("--data-dir="):]; i += 1
        # ... same two-branch pattern repeats verbatim for
        #     --revision, --config, --rules-file, --role, --scope
        #     (six more flags, twelve more branches) ...
        else:
            cleaned.append(tok); i += 1

    # ... config load, revision negotiation (~30 lines) ...

    if resource == "blueprint":
        if cmd == "create":  cmd_create(data_dir, ...)
        elif cmd == "list":  cmd_list(data_dir, ...)
        # ... five more cmd branches ...
        else: emit_error("validation_error", f"unknown command: {cmd}")
    elif resource == "revision":
        # ... same cmd-ladder shape, six more resources ...
\end{lstlisting}
\end{figure}

\subsection{Erosion Sensitivity}\label{sec:erosion-sensitivity}

We vary the high-CC cutoff (8, 10, 12) and the size term (no size term, $\sqrt{\text{SLOC}}$, linear SLOC) around the reported erosion family. Across all nine variants, the result is stable: predictive correlation with next-checkpoint pass rate stays near zero, while predictive correlation with next-checkpoint cost stays positive.

\begin{table}[h]
\centering
\small
\caption{Erosion metric sensitivity sweep.}
\label{tab:appendix-erosion-sensitivity}
\begin{tabular}{@{}lrr@{}}
\toprule
\textbf{Metric} & \textbf{Pass} & \textbf{Cost} \\
\midrule
Reported erosion & $-0.018$ & $0.167$ \\
Best variant (CC > 12 + Linear SLOC) & $-0.136$ & $0.095$ \\
LOC & $-0.212$ & $0.502$ \\
Mean CC & $-0.046$ & $0.218$ \\
Max CC & $-0.117$ & $0.356$ \\
Clone Ratio & $-0.042$ & $0.114$ \\
AST-grep hits / LOC & $-0.052$ & $0.151$ \\
\bottomrule
\end{tabular}
\end{table}

Size-heavy baselines such as LOC and max CC remain stronger raw cost predictors, but the erosion-family conclusion does not depend on the exact threshold or size term.

\needspace{4\baselineskip}
\section{Evaluation Setup}\label{sec:eval-env-details}

\subsection{Environment and Invocation}

\paragraph{Harness version selection.} For each model we report the earliest publicly available harness version that supported that model and could execute the benchmark end-to-end. For older models whose launch-era harness was unavailable or incompatible, we used the nearest later compatible version.

\paragraph{Environment.} The container image installs all languages required by the problem set alongside a shared tooling baseline. We derived this baseline by identifying commands whose absence caused failures across all harnesses; commands that failed on only one harness were excluded to avoid biasing the environment toward a particular agent. Installed packages, shell history, and agent session data reset between checkpoints. The benchmark problems are language-agnostic by design, but current experiments evaluate only the Python track.

\paragraph{Invocation.} Following Terminal Bench~\citep{tb2}, we install Claude Code~\citep{claude_code} and Codex~\citep{codex} directly and invoke each in headless mode. Specific versions appear in \autoref{tab:agent-versions}.

\paragraph{Reasoning effort.} For Codex we set the reasoning effort parameter to \texttt{high}. For Claude Code we configure the thinking-token budget via the environment variable following Anthropic's published mapping.

\subsection{Agent Harness Versions}\label{sec:agent-versions}

\begin{table}[h]
\centering
\caption{Harness version and reasoning effort used for each model in the main evaluation.}
\label{tab:agent-versions}
\small
\begin{tabular}{lllc}
\toprule
Model & Harness & Version & Reasoning \\
\midrule
Opus~4.5      & Claude Code & 2.0.51  & high \\
Opus~4.6      & Claude Code & 2.1.32  & high \\
Sonnet~4.6    & Claude Code & 2.1.44  & high \\
\midrule
GPT~5.2 Codex & Codex CLI & 0.93.0  & high \\
GPT~5.3 Codex & Codex CLI & 0.98.0  & high \\
GPT~5.3 Spark & Codex CLI & 0.100.0 & high \\
GPT~5.4       & Codex CLI & 0.110.0 & high \\
GPT~5.4 Mini  & Codex CLI & 0.110.0 & high \\
\midrule
Composer~2    & Cursor CLI  & 2026.04.13-a9d7fb5 & none \\
GLM~5.1       & Claude Code & 2.1.44  & high \\
Kimi~K2.5     & Claude Code & 2.1.44  & high \\
Kimi~K2.6     & Kimi CLI    & 1.37.0  & high \\
MiniMax~M2.7  & Claude Code & 2.1.44  & high \\
\bottomrule
\end{tabular}
\end{table}

\subsection{\texttt{just-solve} (Baseline)}\label{sec:prompt-intervention-raw}

The minimal baseline used for all primary evaluations. Each agent receives a Jinja template as its system prompt. The \texttt{is\_continuation} flag is false for the first checkpoint of a problem and true for all subsequent checkpoints. The checkpoint specification is injected verbatim via \texttt{spec}.

\begin{lstlisting}[style=specblock, caption={\texttt{just-solve} prompt template}, label={lst:just-solve-prompt}]
Implement a program that 100% solves the specification.
That is all you need to do.
{% if not is_continuation -%}
Use a virtual environment and ensure that a
`requirements.txt` is present with any dependencies
you need to solve the problem.
{% else -%}
Keep using the same virtual environment you started with,
update `requirements.txt` with any new dependencies
you need.
{% endif -%}

Your task is:
{{ spec.strip() }}
\end{lstlisting}

\subsection{\texttt{anti\_slop}}

Explicitly instructs the agent to avoid verbose patterns, defensive over-engineering, and unnecessary abstractions.

\begin{lstlisting}[style=specblock, caption={\texttt{anti\_slop} prompt template}, label={lst:anti-slop-prompt}]
{%- if not is_continuation -%}
You are an exceptional python software engineer and you
to need to implement a spec. Your instructions are:
- Write the python script that satisfies the spec
  completely.
- Use a virtual environment and ensure that a
  `requirements.txt` is present with any dependencies
  you need to solve the problem.
{%- else -%}
You are an exceptional python software engineer and you
to need to implement a spec. You are updating your code
to match an extension of the spec. Here are your
instructions:
- Keep using the same virtual environment you started
  with, update `requirements.txt` with any new
  dependencies you need.
- Focus only on adding in the new features/changes below.
- Make sure you test any examples provided in the task
  description
{%- endif %}
- You ONLY work in this directory.
- Follow best coding practices:
  - Group functions into files based on related
    functionality
  - Keep your code clean
- No god functions/classes.
- Make sure the code is documented appropriately so that
  it is easy to pick up.
- Minimize the following gotchas:
  - Extra defensive checks or try/catch blocks that are
    abnormal.
  - Casts to get around type checking
  - Variables that are only used a single time after
    declaration.
  - Extra comments that a human wouldn't add.
  - Trivial wrappers
  - Heavy nesting
  - If/Else ladders
  - A ton of helper methods

Your task is:
{{ spec.strip() }}
\end{lstlisting}

\subsection{\texttt{plan\_first}}

Requires the agent to plan its approach before writing code.

\begin{lstlisting}[style=specblock, caption={\texttt{plan\_first} prompt template}, label={lst:plan-first-prompt}]
You are an expert programmer and need to implement a task.
{% if not is_continuation -%}
Use a virtual environment and ensure that a
`requirements.txt` is present with any dependencies
you need to solve the problem.
{% else -%}
Keep using the same virtual environment you started with,
update `requirements.txt` with any new dependencies
you need.
{% endif -%}

Here are the steps you should always follow:
1. Before coding plan out what you need to implement.
2. Write the simple solution first.
3. Ensure it is 100% correct and you have covered all
   edge cases.
4. Refactor to ensure the code is high quality.

Here are the basic style rules you must follow:
- Make sure the code is documented appropriately so that
  it is easy to pick up.
- Minimize the following gotchas:
  - Extra defensive checks or try/catch blocks that are
    abnormal.
  - Casts to get around type checking
  - Variables that are only used a single time after
    declaration.
  - Extra comments that a human wouldn't add.
  - Trivial wrappers
  - Heavy nesting
  - If/Else ladders
  - A ton of helper methods
- Follow best coding practices:
  - Group functions into files based on related
    functionality
  - Keep your code clean

Your task is:
{{ spec.strip() }}
\end{lstlisting}

\section{Results Continued}\label{sec:human-repo-panel}

\begin{table}[t]
\centering
\caption{Full set of runs on \scb{} across all harnesses, versions, reasoning levels, and prompt strategies. \textbf{Bold} marks the best value per column.}
\label{tab:performance-appendix}
\footnotesize
\setlength{\tabcolsep}{3pt}
\resizebox{\textwidth}{!}{%
\begin{tabular}{l l l l l | rrrr | rr | rr}
\toprule
 & & & &  & \multicolumn{4}{c|}{Solve Rate (\%)} & \multicolumn{2}{c|}{Cost} & \multicolumn{2}{c}{Quality} \\
\cmidrule(lr){6-9} \cmidrule(lr){10-11} \cmidrule(l){12-13}
Model & Harness & Version & Think & Prompt & Strict & Iso. & Core & Partial & \$/CKPT & Net \$ & Erosion & Verbosity \\
\midrule
Opus 4.5~\citep{claude_opus_45} & Claude Code & 2.0.51 & High & Just Solve & 9.2 & 17.3 & 57.7 & 25.0 & 2.53{\scriptsize$\pm$1.57} & 492.40 & 0.70{\scriptsize$\pm$0.20} & 0.43{\scriptsize$\pm$0.18} \\
\midrule
Opus 4.6~\citep{claude_opus_46} & Claude Code & 2.1.32 & High & Just Solve & 9.7 & 20.9 & \textbf{67.3} & 27.8 & 3.17{\scriptsize$\pm$2.82} & 620.80 & 0.75{\scriptsize$\pm$0.15} & 0.44{\scriptsize$\pm$0.19} \\
\midrule
Opus 4.7~\citep{claude_opus_47} & Claude Code & 2.1.111 & High & Just Solve & 8.2 & 20.9 & 65.8 & 36.1 & 2.17{\scriptsize$\pm$1.83} & 425.34 & 0.76{\scriptsize$\pm$0.17} & 0.48{\scriptsize$\pm$0.21} \\
\midrule
Sonnet 4.6~\citep{claude_sonnet_46} & Claude Code & 2.1.44 & High & Just Solve & 7.1 & 16.8 & 57.7 & 27.8 & 1.96{\scriptsize$\pm$1.97} & 383.83 & 0.75{\scriptsize$\pm$0.20} & 0.44{\scriptsize$\pm$0.19} \\
\midrule
GPT~5.2 Codex~\citep{gpt_52_codex} & Codex CLI & 0.93.0 & High & Just Solve & 9.7 & 21.9 & 56.1 & 36.1 & 0.85{\scriptsize$\pm$0.80} & 167.29 & 0.73{\scriptsize$\pm$0.18} & 0.50{\scriptsize$\pm$0.16} \\
\midrule
GPT~5.3 Codex~\citep{gpt_53_codex} & Codex CLI & 0.98.0 & High & Anti-Slop & 11.2 & 23.5 & 57.1 & 41.7 & 0.73{\scriptsize$\pm$0.50} & 143.04 & 0.42{\scriptsize$\pm$0.20} & 0.31{\scriptsize$\pm$0.12} \\
GPT~5.3 Codex & Codex CLI & 0.98.0 & High & Just Solve & 11.2 & 26.0 & 60.7 & 41.7 & 0.66{\scriptsize$\pm$0.49} & 129.65 & 0.64{\scriptsize$\pm$0.19} & 0.46{\scriptsize$\pm$0.18} \\
GPT~5.3 Codex & Codex CLI & 0.98.0 & High & Plan First & 8.2 & 21.4 & 62.2 & 33.3 & 0.68{\scriptsize$\pm$0.45} & 134.03 & 0.55{\scriptsize$\pm$0.24} & 0.40{\scriptsize$\pm$0.17} \\
\midrule
GPT~5.3 Spark~\citep{gpt_53_spark} & Codex CLI & 0.100.0 & High & Just Solve & 3.1 & 8.2 & 29.1 & 11.1 & \textbf{0.20{\scriptsize$\pm$0.41}} & \textbf{37.17} & 0.68{\scriptsize$\pm$0.19} & 0.48{\scriptsize$\pm$0.16} \\
\midrule
GPT~5.4~\citep{gpt_54} & Codex CLI & 0.110.0 & High & Anti-Slop & 9.2 & 25.5 & 61.2 & 38.9 & 0.90{\scriptsize$\pm$0.78} & 176.46 & 0.28{\scriptsize$\pm$0.16} & 0.24{\scriptsize$\pm$0.09} \\
GPT~5.4 & Codex CLI & 0.110.0 & High & Just Solve & 10.7 & 23.5 & 62.8 & 36.1 & 0.72{\scriptsize$\pm$0.51} & 141.32 & 0.51{\scriptsize$\pm$0.19} & 0.33{\scriptsize$\pm$0.13} \\
GPT~5.4 & Codex CLI & 0.110.0 & High & Plan First & 9.7 & 25.0 & 63.3 & 38.9 & 0.83{\scriptsize$\pm$0.61} & 163.65 & 0.39{\scriptsize$\pm$0.19} & 0.30{\scriptsize$\pm$0.12} \\
\midrule
GPT~5.4 Mini~\citep{gpt_54_mini} & Codex CLI & 0.110.0 & High & Just Solve & 5.1 & 13.8 & 53.1 & 22.2 & 0.45{\scriptsize$\pm$0.35} & 89.03 & 0.66{\scriptsize$\pm$0.17} & 0.42{\scriptsize$\pm$0.12} \\
\midrule
GPT~5.5~\citep{gpt_55} & Codex CLI & 0.124.0 & High & Anti-Slop & 9.2 & 24.0 & 63.3 & 38.9 & 1.68{\scriptsize$\pm$0.83} & 329.65 & \textbf{0.21{\scriptsize$\pm$0.15}} & \textbf{0.21{\scriptsize$\pm$0.08}} \\
GPT~5.5 & Codex CLI & 0.124.0 & High & Just Solve & \textbf{14.8} & \textbf{28.1} & 66.8 & \textbf{50.0} & 1.51{\scriptsize$\pm$0.81} & 295.59 & 0.49{\scriptsize$\pm$0.20} & 0.32{\scriptsize$\pm$0.12} \\
GPT~5.5 & Codex CLI & 0.124.0 & High & Plan First & 8.2 & 24.0 & \textbf{67.3} & 30.6 & 1.61{\scriptsize$\pm$0.72} & 316.26 & 0.29{\scriptsize$\pm$0.18} & 0.25{\scriptsize$\pm$0.12} \\
\midrule
Composer~2~\citep{composer_2} & Cursor CLI & 2026.04.13-a9d7fb5 & None & Just Solve & 6.1 & 16.3 & 52.6 & 13.9 & 0.44{\scriptsize$\pm$0.44} & 86.31 & 0.72{\scriptsize$\pm$0.15} & 0.45{\scriptsize$\pm$0.17} \\
\midrule
GLM~5.1~\citep{glm_51} & Claude Code & 2.1.44 & High & Just Solve & 9.7 & 13.8 & 40.3 & 30.6 & 1.47{\scriptsize$\pm$1.45} & 288.85 & 0.71{\scriptsize$\pm$0.17} & 0.41{\scriptsize$\pm$0.18} \\
GLM~5.1 & OpenCode & 1.4.3 & High & Just Solve & 5.6 & 8.2 & 20.4 & 19.4 & 0.59{\scriptsize$\pm$0.99} & 107.39 & 0.70{\scriptsize$\pm$0.22} & 0.45{\scriptsize$\pm$0.20} \\
\midrule
Kimi~K2.5~\citep{kimi_k25} & OpenCode & 1.4.3 & None & Just Solve & 4.6 & 8.7 & 31.6 & 16.7 & 0.53{\scriptsize$\pm$0.80} & 104.77 & 0.71{\scriptsize$\pm$0.18} & 0.46{\scriptsize$\pm$0.18} \\
Kimi~K2.5 & Claude Code & 2.1.44 & High & Just Solve & 3.6 & 7.1 & 29.6 & 16.7 & 1.07{\scriptsize$\pm$0.96} & 210.36 & 0.71{\scriptsize$\pm$0.18} & 0.43{\scriptsize$\pm$0.19} \\
Kimi~K2.5 & Kimi CLI & 1.37.0 & High & Just Solve & 4.6 & 9.7 & 39.8 & 19.4 & 0.33{\scriptsize$\pm$0.25} & 63.21 & 0.72{\scriptsize$\pm$0.20} & 0.44{\scriptsize$\pm$0.20} \\
\midrule
Kimi~K2.6~\citep{kimi_k26} & Kimi CLI & 1.37.0 & High & Just Solve & 10.7 & 18.9 & 51.0 & 33.3 & 0.74{\scriptsize$\pm$0.64} & 133.18 & 0.76{\scriptsize$\pm$0.20} & 0.51{\scriptsize$\pm$0.22} \\
\midrule
MiniMax~M2.7~\citep{minimax_m27} & Claude Code & 2.1.44 & High & Anti-Slop & 2.6 & 4.1 & 29.1 & 11.1 & 0.33{\scriptsize$\pm$0.21} & 64.92 & 0.51{\scriptsize$\pm$0.22} & 0.38{\scriptsize$\pm$0.15} \\
MiniMax~M2.7 & OpenCode & 1.4.3 & None & Just Solve & 1.5 & 3.6 & 21.9 & 5.6 & 0.27{\scriptsize$\pm$0.36} & 53.36 & 0.75{\scriptsize$\pm$0.20} & 0.45{\scriptsize$\pm$0.20} \\
MiniMax~M2.7 & Claude Code & 2.1.44 & High & Just Solve & 2.0 & 4.1 & 28.1 & 8.3 & 0.34{\scriptsize$\pm$0.23} & 65.79 & 0.73{\scriptsize$\pm$0.16} & 0.47{\scriptsize$\pm$0.17} \\
\bottomrule
\end{tabular}%
}
\end{table}

\subsection{Human Panel construction}

The temporal panel spans \avhPanelN{} maintained Python repositories sampled across web frameworks, scientific computing, infrastructure tools, machine-learning libraries, and command-line utilities. Repositories are stratified by GitHub star tier: Hobby ($<100\bigstar$, $n{=}\avhHobbyN$), Niche ($100$--$1\text{k}\bigstar$, $n{=}\avhNicheN$), Established ($1$--$10\text{k}\bigstar$, $n{=}\avhEstabN$), and Major ($\geq10\text{k}\bigstar$, $n{=}\avhMajorN$). For each repository we sample up to 30 source-touching commits across its history, totaling \avhTemporalCkpts{} checkpoints between \avhPanelStartYear{} and \avhPanelEndYear{}. Checkpoints with empty source trees or failed metric runs are dropped before analysis. Each repository's most recent checkpoint serves as its HEAD snapshot for the calibration band reported in \autoref{sec:agent-vs-human}. \autoref{tab:agent-vs-human} reports HEAD verbosity and erosion by tier and against agent checkpoints.

\paragraph{Pre- vs.\ post-ChatGPT temporal check.} The panel straddles 2022, so post-ChatGPT commits could in principle include LLM-assisted contributions. Pre-ChatGPT checkpoints ($n{=}\avhTemporalPreN$) have a median verbosity of \avhTemporalPreVerbMedian{} versus \avhTemporalPostVerbMedian{} for post-ChatGPT checkpoints ($n{=}\avhTemporalPostN$). The shift is real but small relative to the agent gap (mean verbosity \avhAgentVerbMean{}). The same comparison anchored at January 2024 yields medians of \avhAgentEraPreVerbMedian{} versus \avhAgentEraPostVerbMedian{} for verbosity and \avhAgentEraPreErosMedian{} versus \avhAgentEraPostErosMedian{} for erosion.

\paragraph{Within-repository agent-era shift.} A cross-section across the 2024 boundary mixes new repositories in with old ones. Restricting to the \avhAgentEraBothN{} repositories with at least three checkpoints in each era, the median within-repository verbosity shift is $+\avhAgentEraVerbShiftMedian$ and the erosion shift is $+\avhAgentEraErosShiftMedian$. \avhAgentEraVerbHigherPct{}\% of these repositories are more verbose post-2024 and \avhAgentEraErosHigherPct{}\% are more eroded. The signal is consistent with a small uptick in human-authored slop after agent-assisted contributions became common, but it is dwarfed by the agent-side trajectory growth reported in \autoref{sec:agent-vs-human}.

\paragraph{First-to-last growth.} For repositories with at least five sampled commits, \avhTemporalVerbRising{} of \avhTemporalVerbTotal{} ($\avhTemporalVerbRisingPct{}\%$) end with higher verbosity than they started, with median first-to-last growth of \avhTemporalVerbGrowthMedian{}\%. Erosion rises in \avhTemporalErosRising{} of \avhTemporalErosTotal{} ($\avhTemporalErosRisingPct{}\%$) repositories. Per-checkpoint slope is \avhHumanVerbSlopeMedian{} for verbosity and \avhHumanErosSlopeMedian{} for erosion across the human panel, against \avhAgentVerbSlopeMedian{} and \avhAgentErosSlopeMedian{} for agent trajectories.

\paragraph{Notable outliers.} \texttt{huggingface/transformers} (verbosity \avhTransformersVerb{}, erosion \avhTransformersErosion{}) and \texttt{fastapi/fastapi} (\avhFastapiVerb{}, \avhFastapiErosion{}) are the most prominent verbosity outliers in the panel, both above the agent verbosity mean. The high transformers verbosity is driven by per-architecture file duplication (one model implementation per family). At the other end, \texttt{pallets/flask} (\avhFlaskVerb{}, \avhFlaskErosion{}) and \texttt{scikit-learn/scikit-learn} (\avhSklearnVerb{}, \avhSklearnErosion{}) are large, mature projects with low verbosity and middling erosion. \texttt{scipy/scipy} (\avhScipyVerb{}, \avhScipyErosion{}) sits at the high end of human erosion, still well below the agent mean.

\section{Extended Related Work}\label{sec:appendix-extended-related-work}

\paragraph{Quality degradation under multi-turn coding.} LLM-generated code degrades under repeated modification. Code converges toward structural attractors~\citep{iterative-readability}, quality diverges across agents and runs~\citep{chen2025-patch-quality, santos2025-software-aging, bohr2025-show-and-tell}, and refinement introduces defects that correctness testing does not catch~\citep{scaffold-cegis, diff-fuzzing-refactoring}. Interaction failure modes compound these effects~\citep{interaction-smells, overcorrection-reviewers, more-rounds-more-noise}; agent-generated code has become a practical integration burden~\citep{nakashima2026-agentic-prs, vibe-coding-overhead}.

\paragraph{Code quality metrics.} LLM code has higher cyclomatic complexity per line than human-written code~\citep{dou2024-whats-wrong-llm-code}, though aggregate CC can fall while vulnerabilities rise~\citep{cotroneo2025-human-vs-ai}. \citet{abbassi2025-taxonomy-inefficiencies} extend the code smell taxonomy to LLM outputs, finding redundant steps, duplication, and unnecessary conditionals most prevalent. Code smells are well-studied in software engineering~\citep{fowler1999-refactoring, lacerda2020-code-smells-tertiary}; software aging~\citep{parnas1994-software-aging} and technical debt~\citep{cunningham1992-technical-debt} describe how structural degradation accumulates under modification, with \citet{li2022-architecture-erosion} and \citet{le2021-architectural-decay} documenting this in long-lived systems.

\paragraph{Benchmark landscape.} Beyond the paradigm established by \citet{swe-bench}, the evaluation landscape spans three waves. The first broadens language and domain coverage through extensions~\citep{swe-bench-plus, multi-swe-bench, rust-swe-bench, swe-rebench-v2, swe-polybench}. Instruction-following benchmarks evaluate compliance across conversation turns~\citep{codeif-bench, multicodeif} but assess each response independently. A second wave targets feature-level development from existing repositories~\citep{fea-bench, featurebench}. A third wave builds entire projects from scratch: \citet{commit0} generate libraries from specifications with interactive test feedback, several benchmarks construct full repositories from natural-language requirements~\citep{projdevbench, e2edevbench, nl2repo-bench, swe-agi, projecteval}, and \citet{longcli-bench} spans from-scratch construction through refactoring. Tasks grow larger and harder across all three waves; evaluation in each case is a single artifact assessed against a fixed specification.

\end{document}